\documentclass[journal]{IEEEtran}
\usepackage{latexsym}
\usepackage{graphicx}
\usepackage{amsfonts,amssymb,amsmath}
\usepackage{mathtools, cuted}
\usepackage{hyperref}

\usepackage[T1]{fontenc}
\usepackage{cite}
\usepackage{subcaption}
\usepackage{comment}

\usepackage{amsthm}
\usepackage{overpic}
\usepackage{steinmetz}
\usepackage{array}
\usepackage{url}
\usepackage{color}

\usepackage{algorithmicx}
\usepackage{algcompatible}
\usepackage[ruled,linesnumbered,noend]{algorithm2e}

\theoremstyle{plain}

\newtheorem{definition}{Definition}

\newtheorem{remark}{Remark}
\newtheorem{proposition}{Proposition}

\newcommand{\vect}[1]{\mathbf{#1}}

\def\diag{\mathrm{diag}}

\def\Htran{\mbox{\tiny $\mathrm{H}$}}
\def\Ttran{\mbox{\tiny $\mathrm{T}$}}
\def\CN{\mathcal{N}_{\mathbb{C}}} %Complex Gaussian
\def\mod{\mathrm{mod}}

\IEEEoverridecommandlockouts

\begin{document}

\title{On Broad-Beam Reflection for Dual-Polarized RIS-Assisted MIMO Systems }

\author{Parisa Ramezani, Maksym A. Girnyk, and  Emil Bj\"{o}rnson

\thanks{%
P. Ramezani and E. Bj\"{o}rnson
are with the Department of Computer Science, KTH Royal Institute of Technology, Stockholm, Sweden (email:\{parram,emilbjo\}@kth.se). They are supported by Grant 2019-05068 from the Swedish Research Council and the FFL18-0277 grant from the Swedish Foundation for Strategic Research. M. A. Girnyk is with Ericsson Networks, Stockholm, Sweden (e-mail: max.girnyk@ericsson.com).\\
A preliminary version of this paper \cite{Parisa2023Asilomar} has been presented at the IEEE Asilomar Conference on Signals, Systems, and Computers, 2023.}}

\maketitle

\begin{abstract}
  The use of a reconfigurable intelligent surface (RIS) for aiding user-specific transmission has been widely explored. However, little attention has been devoted to utilizing RIS for assisting cell-specific transmission, where the RIS needs to reflect signals in a broad angular range. Furthermore, although modern communication systems operate in two polarizations, the majority of the works on RIS consider a uni-polarized surface, only reflecting the signals in one polarization. 
  To fill these gaps, we study a downlink broadcasting scenario where a base station (BS) sends a cell-specific signal to all the users residing at unknown locations with the assistance of a dual-polarized RIS. We utilize the duality between the auto-correlation function and power spectrum in the space/spatial frequency domain to design configurations for broad-beam reflection. We first consider a free-space line-of-sight BS-RIS channel and show that the RIS configuration matrices must form a Golay complementary array pair for broad-beam radiation. We also present how to form Golay complementary array pairs based on known Golay complementary sequence pairs.  We then consider an arbitrary BS-RIS channel and propose an algorithm based on stochastic optimization to find RIS configurations that produce a practically broad beam by relaxing the requirement on uniform broadness. Numerical simulations are finally conducted to corroborate the analyses and evaluate the performance.    
\end{abstract}
\begin{IEEEkeywords}
Reconfigurable intelligent surface, broad beamforming, dual-polarized communication, Golay complementary pairs, power-domain array factor.
\end{IEEEkeywords}

\section{Introduction}
\begin{figure}
	\centering
	\begin{subfigure}{0.5\columnwidth}
		\centering
		\includegraphics[width=0.95\textwidth]{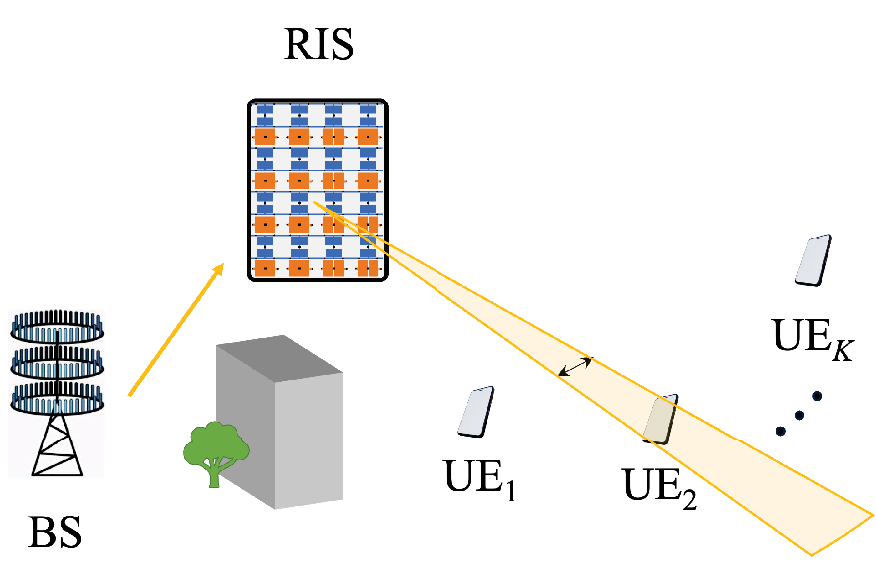}
		\caption{UE-specific transmission.}
		\label{fig:ue_spec_ris}
	\end{subfigure}%
	\begin{subfigure}{0.5\columnwidth}
		\centering
		\includegraphics[width=0.95\textwidth,trim={0 0.6cm 1.5cm 0},clip]{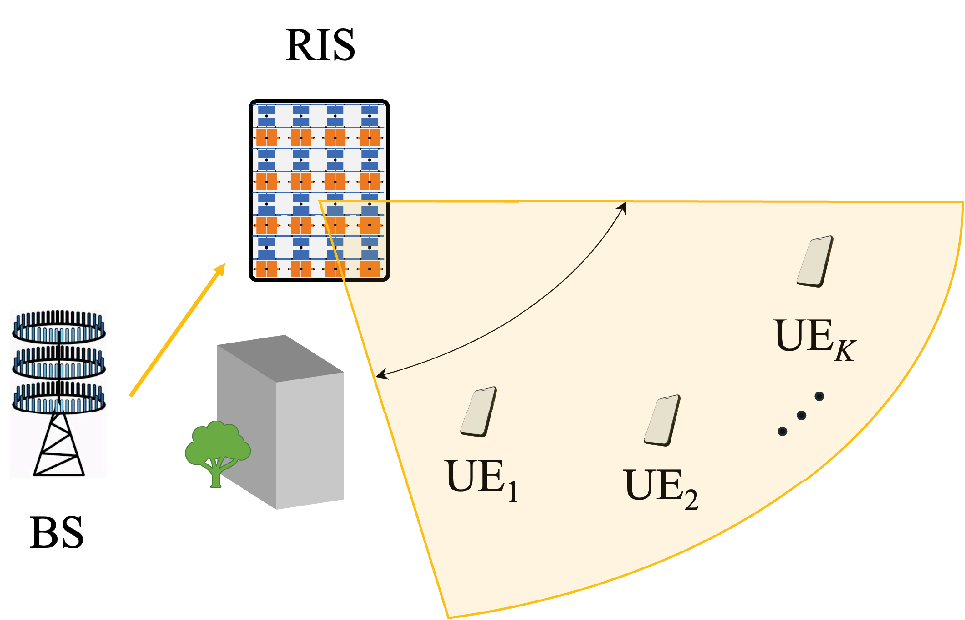}
		\caption{Sector-specific transmission.}
		\label{fig:cell_spec_ris}
	\end{subfigure}
	\caption{Various types of RIS-aided transmission.}
\label{fig:ris_transmission}
\vspace{-0.5cm}
\end{figure}

In recent years, we have witnessed an upsurge of interest in the development of reconfigurable intelligent surface (RIS)-assisted communication, where RIS has been utilized to enhance the performance of various systems, e.g., cell-free massive multiple-input multiple-output (MIMO) systems \cite{Shi2024RIS}, wireless powered communication networks \cite{Lyu2021Optimized}, multi-operator networks \cite{Doga2024Combating}, etc. An RIS is an auxiliary network entity, engineered to control the propagation of electromagnetic waves intelligently \cite{DiRenzo2020Smart}. Despite the extensive research conducted in this area, some fundamental problems have remained untouched. Particularly, the majority of prior works have considered user-specific RIS-aided communication depicted in Fig.~\ref{fig:ue_spec_ris}. In such a setting, the phase shifts of RIS elements are designed in a way to form a narrow beam towards one (or a few) user(s) based on user-specific channel state information (CSI) \cite{Guo2020, Dong2020,Lyu2021Optimized}.  
However, in many practical situations, numerous users could be spread over a wide angular sector and possibly at unknown locations, while requiring simultaneous service, as depicted in Fig.~\ref{fig:cell_spec_ris}. A typical example is the broadcasting phase, where fundamental network and communication parameters are exchanged between the base station (BS) and any potential user that resides in the coverage area. Specifically, at the beginning of communication when the BS does not have any information about the users' locations, it needs to announce its existence and characteristics by broadcasting common messages over its entire coverage area to tell prospective users how to connect to the BS.  One way to reach this objective is via beam sweeping, where the RIS generates narrow beams directed toward all potential angular directions. This approach, however, is resource-intensive. Alternatively, the RIS can be configured to reflect the signal as a single broad beam, uniformly covering a broad angular range, thereby offering a more efficient use of resources. 
Broad beamforming is also useful for supporting user mobility as it ensures continuous service to the mobile user as long as the user remains within the coverage area of the RIS. The resulting SNR in this case will be lower than with user-specific beamforming, but no pilot signaling is required to track the user and continuously change the beam direction.

Considering the need for broadcasting cell-specific signals, it is important to devise methods for producing broad beams from the RIS to uniformly cover all angular areas of interest. Broad-beam design for RIS is more challenging than that for active antenna arrays because the latter can use any beamforming coefficients to produce a broad beam, while the RIS has to reflect the distorted signal of the transmitter as a broad beam by compensating for the transmitter-RIS channel and it can only perform phase shifting due to passiveness.

Recently, beam broadening methods have been proposed in \cite{Haiquan2021,He2023BroadCoverage,AlHajj2023,Haghshenas2024Parametric,Lin2024} for partially widening the beam reflected by the RIS. In particular, \cite{Haiquan2021} considers an aerial RIS-assisted communication and develops a beam broadening and flattening technique for producing beams with adjustable beamwidths to cover a target area. They partition the RIS into several sub-arrays and design the phase shifts of the sub-arrays such that a wide beam is generated by combining the beams produced by all of them.  Reference \cite{He2023BroadCoverage} aims to achieve a quasi-static broad coverage by minimizing the difference between
a predefined pattern and the RIS-reflected power pattern using statistical CSI. The authors in \cite{AlHajj2023} propose an optimization framework based on the genetic algorithm to design a codebook of wide beams, aiming to minimize the beam ripple and sidelobe level. Reference \cite{Haghshenas2024Parametric} first designs a novel low-dimension codebook of RIS configurations and then 
proposes to split the RIS into multiple sub-arrays such that the number of sub-arrays approximately equals the codebook size of the sub-arrays. By configuring each sub-array with one of the columns of the designed codebook, the RIS is shown to produce a nearly isotropic beam that covers all angular directions. 
In \cite{Lin2024}, the authors design a broad reflected beam with equal received power in all angular directions of a predefined sector, while allowing for some fluctuations in the generated broad-beam pattern.

All of the above works on broad-beam design for RIS considered a uni-polarized surface.  However, modern communication systems (e.g., 5G) operate over two orthogonal polarizations, typically vertical and horizontal or slanted $\pm 45$ degrees \cite{Girnyk2023Broad}.  Dual-polarized communication offers several advantages over uni-polarized communication, including doubling the capacity without requiring additional spectrum and enhancing reliability by improving resilience to signal degradation due to environmental factors such as multi-path fading and polarization mismatch. Therefore, if an RIS is to be seamlessly integrated into future communication systems, it must be equipped with two polarizations to receive signals on both polarizations and control the reflections independently \cite{Chen2021}. A uni-polarized RIS is inefficient as it is only capable of controlling the reflection of signals in one polarization.

 Similar to uni-polarized systems, dual-polarized systems must switch between common information transmission and specific data transmission. While there have been some recent works on rate evaluations for dual-polarized RIS-assisted mobile communications and RIS-enabled holographic MIMO systems \cite{Han2022Dual}, \cite{Zeng2024Dual}, designing effective broadcasting mechanisms in such systems has been largely overlooked. 
To ensure that the benefits of dual-polarized communication are fully reaped, we need new broad-beam designs so that the BS can effectively reach out to users before commencing the communication. This paper aims to fill the aforementioned gap by designing broad RIS-reflected beams to support common information broadcast from the BS in dual-polarized RIS-assisted systems.  
   
A broad beam is specified as a beam whose power-domain array factor is (approximately) constant over all angular directions of interest, resulting in consistent signal strength across the coverage area.  In essence, the radiation pattern of a broad beam resembles that of a single array element, but is shifted upwards by the power-domain array factor \cite{Girnyk2023Broad}. In dual-polarized systems, the polarization degree of freedom can be utilized to produce broad radiation patterns \cite{Max2022,Max2021,Li2021Golay} that are not possible with a single polarization. The idea is to design complementary radiation patterns for the two polarizations such that the combination of the per-polarization radiated beams results in a broad radiation pattern; that is, one polarization has a peak in the array factor where the other polarization has a null, and vice versa. 
In the context of RIS-assisted communication, the recent work in  \cite{ParisaBroadBeamLetter} considers a dual-polarized uniform linear array (ULA)-type RIS and designs the per-polarization RIS configuration vectors to obtain a uniformly broad reflected beam, assuming a line-of-sight (LoS) channel between the BS and the RIS. 

The present paper considers the more practical uniform planar array (UPA) structure for the dual-polarized RIS and aims to design the per-polarization phase configurations such that the power-domain array factor of the surface becomes spatially flat and the radiation pattern of the RIS becomes an elevated version of a single element radiation pattern,  thus enabling the RIS to radiate a broad beam.  
We first consider a LoS channel scenario between the BS and the RIS and show that a uniformly broad beam can be achieved in this case if the RIS configuration matrices form a Golay complementary array pair \cite{Golay1951,Sivaswamy1978}. 
 We then investigate the general case with an arbitrary but known channel between the BS and the RIS and propose a stochastic optimization algorithm for designing the RIS configurations for broad-beam reflection. This paper is an extended and refined version of our conference paper \cite{Parisa2023Asilomar} in which we presented some preliminary analyses and results on broad-beam design for dual-polarized RIS considering a purely LoS channel between the BS and the RIS.

The contributions of this paper are summarized as follows:
\begin{itemize}
    \item We consider a dual-polarized RIS-assisted communication system, aiming to design the RIS configurations of the two polarizations such that the RIS radiates a broad beam to support common information transmission to users at unknown locations. The radiated beam must uniformly cover all target azimuth and elevation angles, making the design of UPA-type RIS configurations more challenging compared to the ULA case in \cite{ParisaBroadBeamLetter} where only azimuth angles were considered. 
    \item {We introduce Golay complementary pairs and show how their unique property in having complementary auto-correlation functions (ACFs) results in a constant sum power spectrum. We then prove that when the channel between the BS and the RIS is purely LoS, a uniformly broad beam is only achievable if the RIS configuration matrices are set to be a Golay complementary array pair. We present methods for constructing Golay complementary array pairs and further elaborate on how small-size Golay array pairs can be utilized to construct larger ones. }
    
    \item We then address a more practical scenario with an arbitrary channel between the BS and RIS with non-LoS (NLoS) components. In this setting, Golay complementary pairs are unsuitable for broad beamforming due to fluctuations in the BS-RIS link amplitude. To address this, we develop a heuristic algorithm using stochastic optimization to design broad-beam-producing RIS configurations. This algorithm permits a controlled level of imperfection in the sum ACF, specified by the threshold $\epsilon$.
    
    \item We numerically assess the performance of our broad-beam designs. We show that using Golay complementary configurations, the RIS generates a uniform broad beam and retains the shape of a single element's beam. Further, we demonstrate that the $\epsilon$-complementary algorithm enables the RIS to achieve near-uniform angular coverage. We also evaluate end-user performance by examining the achievable spectral efficiency (SE). Simulations show that our broad beamforming schemes effectively serve users in unknown locations with significantly higher minimum SE than benchmarks, highlighting the advantage of a dual-polarized RIS over a uni-polarized one.

\end{itemize}

The remainder of this paper is organized as follows: Section~\ref{sec:sys_mod} presents the system model of a dual-polarized RIS-assisted communication with a LoS channel between the BS and the RIS. In Section~\ref{sec:Golay_pairs}, we present our broad-beam design approach using the Golay complementary pairs and Section~\ref{sec:Golay_construction} explains how Golay complementary array pairs can be constructed. We extend the system model to the general arbitrary channel between the BS and the RIS in Section~\ref{sec:epsilon_comp} and present the $\epsilon$-complementary algorithm for designing practically broad beams in this case. Numerical results are provided in Section~\ref{sec:results}, while Section~\ref{sec:conclusions} concludes the paper and provides future outlook on possible research in this area. 

\textbf{Notations:} Scalars are denoted by italic letters, vectors and matrices are denoted by bold-face lower-case and uppercase letters, respectively. $(\cdot)^*$, $(\cdot)^{\Ttran}$, and $(\cdot)^{\Htran}$ indicate the conjugate, transpose, and conjugate transpose, respectively. $\mathbb{C}$ denotes the set of complex numbers. $\otimes$ and $\odot$ represent the Kronecker product and Hadamard product, respectively. $\diag(a_1,\ldots,a_N)$ is a diagonal matrix having $a_1,\ldots,a_N$ as its diagonal elements. $|\cdot|$ denotes the absolute/magnitude value, $[\vect{a}]_i$ is the $i$th entry of the vector $\vect{a}$, and $[\vect{A}]_{i,j}$ is the entry in the $i$th row and $j$th column of matrix $\vect{A}$. Moreover, $\delta[\cdot]$ denotes the Kronecker delta function.

\section{System Model}
\label{sec:sys_mod}
\begin{figure}[t]
    \centering
    \includegraphics[width = 0.55\columnwidth]{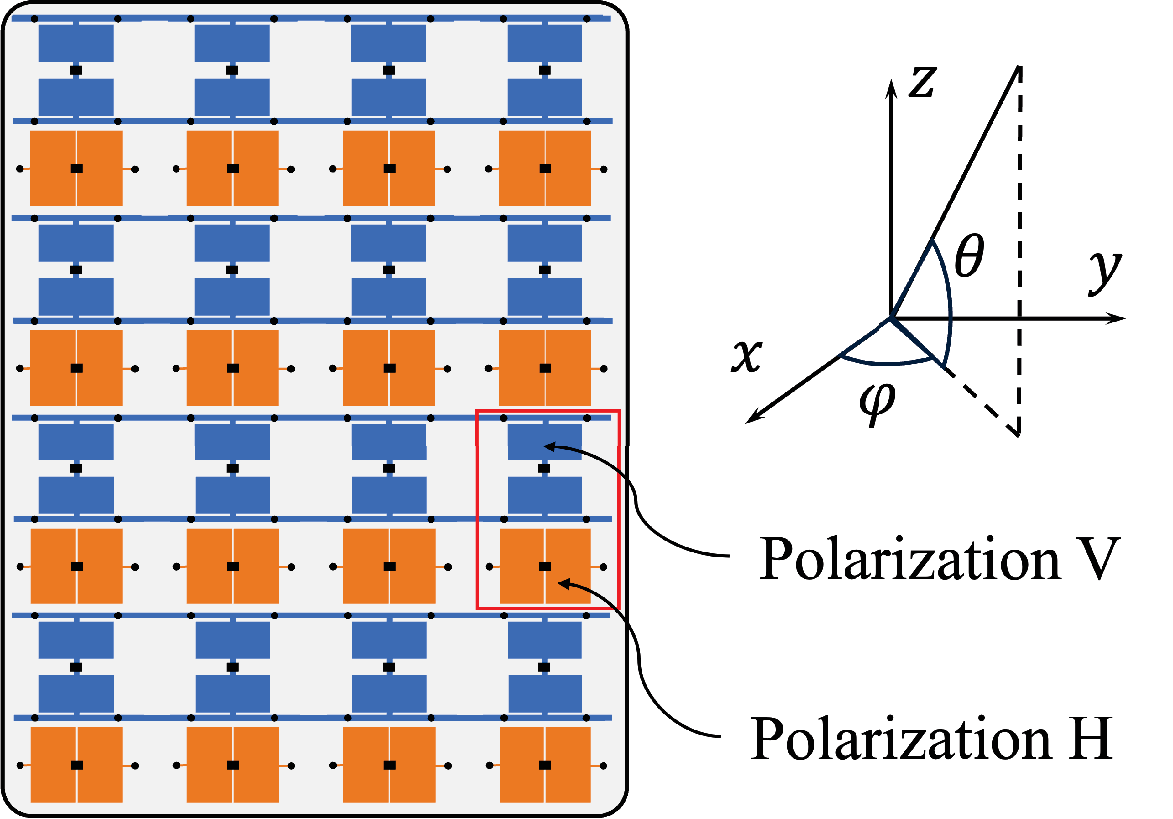}
 \caption{A dual-polarized RIS consisting of meta-atoms with two orthogonal separately controllable polarizations \cite{Chen2021,ke2021linear}.} 
    \label{fig:dp-RIS}
\end{figure}

We consider a dual-polarized RIS-assisted system with a BS communicating with a population of users via the RIS. In particular, we study a broadcast communication scenario where the BS intends to transmit a common signal to all the users who reside in a wide angular sector, and possibly at unknown locations.
The direct links between the BS and users are assumed to be blocked and communication can only take place through the RIS.\footnote{In practice, there will surely be non-zero direct links between the BS and some prospective user locations. Nevertheless, it is desirable for the RIS to act as an isotropic reflector to maximize its spread of the broadcast signals so they can reach potential users in any direction without relying on any information about the availability of the direct links on every location in the coverage area.}
The BS is equipped with $M$ dual-polarized antennas in the form of a ULA and each user has one dual-polarized antenna. The RIS is assumed to have $2N$ elements, out of which, $N$ elements have $\mathrm{H}$ polarization and the other $N$ elements have $\mathrm{V}$ polarization.\footnote{$\mathrm{H}$ and $\mathrm{V}$ refer to horizontal and vertical polarizations, but the results of this paper hold for any pair of orthogonal polarizations.} The elements are arranged in a way that the polarization changes between different RIS rows, as depicted in Fig.~\ref{fig:dp-RIS}.\footnote{The elements in different polarizations can also be arranged such that the polarization changes between different RIS columns \cite{Chen2021}. However, we opted for the design in Fig.~\ref{fig:dp-RIS} because more elements in one polarization in the horizontal dimension provides a better horizontal resolution \cite[Chapter 4]{bjornson2024introduction} which is a desirable feature when the RIS is used for user-specific beamforming.} 

Suppose $s$ is the signal transmitted by the BS. We will analyze the received signal at an arbitrary user in the coverage area and later ensure that it achieves the same SNR regardless of its angular location, with only a scaling factor related to the RIS element gain in each specific direction. We omit the user index for notational convenience. The received signal in polarization $\mathrm{p} \in \{\mathrm{H,V}\}$  at an arbitrary considered user in the azimuth angle $\varphi$ and elevation angle $\theta$ from the RIS can be expressed as \cite [Chapter 9]{bjornson2024introduction} 
\begin{equation}
\label{eq:received_signal}
r_{\mathrm{p}} = \sqrt{P_{\mathrm{T}} \Tilde{\beta}\beta G_{\mathrm{R},0}(\varphi,\theta)}\vect{a}_{\mathrm{p}}^{\Ttran} (\varphi,\theta) \boldsymbol{\Phi}_{\mathrm{p}} \vect{H}_{\mathrm{p}} \vect{w}_{\mathrm{p}} s + n_{\mathrm{p}}, 
\end{equation}where $P_{\mathrm{T}}$ is the BS transmit power, $\Tilde{\beta}$ and $\beta$ are respectively the path-loss from the BS to the RIS and from the RIS to the potential user, $G_{\mathrm{R},0}(\varphi,\theta)$ denotes the RIS element gain,
  $\vect{H}_{\mathrm{p}} \in \mathbb{C}^{N \times M}$ represents the channel between the BS and the RIS, $\vect{w}_{\mathrm{p}} \in \mathbb{C}^M$ is the beamforming vector applied to $s$ by the BS, and $n_{\mathrm{p}}$ is the additive white Gaussian noise at the user with power $\sigma^2$. 
  $\boldsymbol{\Phi}_{\mathrm{p}}$ is the diagonal phase configuration matrix of the RIS with $\boldsymbol{\Phi}_{\mathrm{p}} = \diag{\left(e^{j\omega_{\mathrm{p},1}},\ldots,e^{j\omega_{\mathrm{p},N}}\right)}$, where $\omega_{\mathrm{p},n}$ denotes the phase shift applied by the $n$th RIS element in polarization $\mathrm{p}$ to the incident signal. Furthermore, $\vect{a}_{\mathrm{p}}(\varphi,\theta)$ is the RIS array response vector given by \cite{Swindlehurst2022}
\begin{equation}
    \vect{a}_{\mathrm{p}}(\varphi,\theta)= \vect{a}_{\mathrm{p}}^z (\theta) \otimes \vect{a}_{\mathrm{p}}^y(\varphi,\theta),~~\mathrm{p}\in\{\mathrm{H,V}\},
\end{equation}with $\vect{a}_{\mathrm{p}}^y(\varphi, \theta)$ and $\vect{a}_{\mathrm{p}}^z (\theta)$ denoting the RIS response vectors in $y$ and $z$ dimensions. The number of elements in each row and column of the RIS are respectively denoted by $N_y$ and $N_z$ such that $2N=N_y N_z$, where $N_z$ is assumed to be an even number so that we have the same number of elements in both polarizations, while $N_y$ can be even or odd. According to Fig.~\ref{fig:dp-RIS}, we have 
\begin{align}
   \vect{a}_{\mathrm{V}}^y (\varphi,\theta)&=  \big[1, e^{-j \psi_y },\ldots,e^{-j(N_y -1)\psi_y} \big]^{\Ttran}, \\
   \vect{a}_{\mathrm{V}}^z (\theta) &=  \Big[1, e^{-j2\psi_z},\ldots,e^{-j2(\widetilde{N}_z -1)\psi_z} \Big]^{\Ttran},
\end{align}for the $\mathrm{V}$ polarization, where $\widetilde{N}_z = N_z/2$ and the relative phase shifts are given by 
\begin{equation}
   \label{eq:psi}
   \psi_y = \frac{2\pi}{\lambda}\Delta_y \sin(\varphi) \cos(\theta),\quad\psi_z = \frac{2\pi}{\lambda} \Delta_z \sin(\theta),
\end{equation}in which $\lambda$ is the wavelength of the transmitted signal, and $\Delta_y$ and $\Delta_z$ denote the inter-element spacings between adjacent elements in respective dimensions.
For polarization $\mathrm{H}$, we have $\vect{a}_{\mathrm{H}}^y(\varphi,\theta) = \vect{a}_{\mathrm{V}}^y (\varphi, \theta)$ and $\vect{a}_{\mathrm{H}}^z (\theta)= e^{-j\psi_z} \vect{a}_{\mathrm{V}}^z(\theta)$ due to the vertical shift by one row, which results in 
\begin{equation}
\vect{a}_{\mathrm{H}}(\varphi,\theta) = e^{-j\psi_z} \vect{a}_{\mathrm{V}} (\varphi,\theta).
\end{equation}

  Assuming an ideal free-space LoS channel between the BS and the RIS, $\vect{H}_{\mathrm{p}}$ is modeled as
  \begin{equation}
    \vect{H}_{\mathrm{p}} = \sqrt{G_{\mathrm{R,0}}(\Tilde{\varphi},\Tilde{\theta}) G_{\mathrm{B,0}} (\vartheta)} \vect{a}_{\mathrm{p}}(\Tilde{\varphi},\Tilde{\theta}) \vect{b}^{\Ttran}(\vartheta),  
  \end{equation}where  $G_{\mathrm{B},0}(\cdot)$ is the gain of one BS antenna, and $\Tilde{\varphi}$ and $\Tilde{\theta}$ are the azimuth and elevation angle-of-arrival (AoA) to the RIS, respectively. Furthermore, $\vartheta$ is the angle-of-departure (AoD) from the BS and $\vect{b}(\vartheta)$ is the BS array response vector given by 
  \begin{equation}
    \vect{b}(\vartheta) = \left[1,e^{-j\psi_{\mathrm{B}}},\ldots, e^{-j(M-1)\psi_{\mathrm{B}}} \right]^{\Ttran},  
  \end{equation}where $\psi_{\mathrm{B}} = (2\pi/\lambda)\Delta_{\mathrm{B}} \sin(\vartheta)$  with $\Delta_{\mathrm{B}}$ denoting the spacing between BS antennas. Applying maximum ratio transmission at the BS, the beamforming vectors are set as $\vect{w}_{\mathrm{p}} = \vect{b}^*(\vartheta)/\|\vect{b}(\vartheta)\|,~\mathrm{p} \in \{\mathrm{H},\mathrm{V}\}$. The received signal in \eqref{eq:received_signal} can now be re-expressed as  
\begin{equation}
\begin{aligned}
\label{eq:received_signal2}
   r_{\mathrm{p}} =&\; \sqrt{M P_{\mathrm{T}} \beta \Tilde{\beta}  G_{\mathrm{B,}0}(\vartheta)G_{\mathrm{R,}0}(\Tilde{\varphi},\Tilde{\theta})G_{\mathrm{R,}0}(\varphi,\theta)}\\ &\times \boldsymbol{\phi}_{\mathrm{p}}^{\Ttran} \left( \vect{a}_{\mathrm{p}} (\varphi,\theta) \odot \vect{a}_{\mathrm{p}} (\tilde{\varphi},\tilde{\theta})  \right)s + n_{\mathrm{p}}, 
   \end{aligned}
\end{equation}where $\boldsymbol{\phi}_{\mathrm{p}} = [e^{j\omega_{\mathrm{p},1}},\ldots,e^{j\omega_{\mathrm{p},N}}]^{\Ttran}$. 
After applying maximum ratio combining over the two polarizations, the SNR at the potential user is obtained as 
\begin{equation}
\label{eq:SNR_LoS}
        \mathrm{SNR} = \gamma(\varphi,\theta) \sum_{\mathrm{p}  \in \{\mathrm{H,V}\}}\left|\boldsymbol{\phi}_{\mathrm{p}}^{\Ttran}\big(\vect{a}_{\mathrm{p}}(\varphi,\theta) \odot \vect{a}_{\mathrm{p}}(\Tilde{\varphi},\Tilde{\theta})\big)\right|^2,
\end{equation}where 
\begin{equation}
\gamma(\varphi,\theta) = \frac{M P_{\mathrm{T}} \Tilde{\beta}   \beta G_{\mathrm{B,}0}(\vartheta)G_{\mathrm{R,}0}(\Tilde{\varphi},\Tilde{\theta})G_{\mathrm{R,}0}(\varphi,\theta)}{\sigma^2}. 
\end{equation}We call the term  
\begin{equation}
\label{eq:PDAF}
A(\varphi,\theta) = \left|\boldsymbol{\phi}_{\mathrm{H}}^{\Ttran}\big(\hat{\vect{a}}_{\mathrm{H}}(\varphi,\theta)\big)\right|^2 + \left|\boldsymbol{\phi}_{\mathrm{V}}^{\Ttran}\big( \hat{\vect{a}}_{\mathrm{V}}(\varphi,\theta)\big) \right|^2
\end{equation} the \textit{power-domain array factor} with $\hat{\vect{a}}_{\mathrm{p}}(\varphi,\theta) = \vect{a}_{\mathrm{p}}(\varphi,\theta) \odot \vect{a}_{\mathrm{p}}(\tilde{\varphi},\tilde{\theta})$ being the equivalent array response vector of the RIS in polarization $\mathrm{p}$. Note that $\tilde{\varphi}$ and $\tilde{\theta}$ are assumed to be known due to the fixed location of the BS and the RIS. The aim of this paper is to design the RIS configurations in polarizations $\mathrm{H}$ and $\mathrm{V}$ such that the beam re-radiated from the RIS covers all angular directions, which is achieved when the power-domain array factor is constant regardless of where the user is. The upcoming sections shed light on the details of the proposed designs.

\section{Broad-Beam Design Using Golay Complementary Pairs}
\label{sec:Golay_pairs}

A uniformly broad beam is referred to as a beam whose power-domain array factor is spatially flat over all possible observation angles $(\varphi,\theta)$, i.e.,
\begin{equation}
    \label{eq:PDAF_constant}
    A(\varphi,\theta) = c,~~\varphi \in \left[-\frac{\pi}{2},\frac{\pi}{2} \right],~\theta \in \left[-\frac{\pi}{2},\frac{\pi}{2} \right],
\end{equation}
where $c$ is a constant.

\subsection{Golay Complementary Pairs}
To design the RIS configurations for the $\mathrm{H}$ and $\mathrm{V}$ polarizations, we first describe the concept of Golay complementary sequence pairs, which was introduced by Golay in \cite{Golay1951}.

\begin{definition}[Golay complementary sequence pair]
\label{def:Golay-pair}
    Unimodular sequences $\vect{u} \in \mathbb{C}^N$ and $\tilde{\vect{u}} \in \mathbb{C}^N$ form a Golay complementary sequence pair if and only if 
    \begin{align}
    \label{eqn:conditionGolay}
    	R_{\vect{u}}[\xi] + R_{\tilde{\vect{u}}}[\xi] &= 2N \delta [\xi],
    \end{align}where $R_{\vect{u}} [\xi]$ indicates the ACF of $\vect{u}$ and is given by 
    \begin{small}
    \begin{align}
    \label{eq:ACF}
    	R_{\vect{u}}[\xi] = \left\{ 
    	\begin{aligned} 
    	&\sum\limits_{n=1}^{N-\xi} [\vect{u}]_n [\vect{u}]_{n+\xi}^*, 
    	 & & \xi=0,\ldots,N-1,\\
    	&\sum\limits_{n=1}^{N+\xi} [\vect{u}]_{n-\xi} [\vect{u}]_{n}^*, 
    	 & &\xi = -N+1,\ldots,-1,\\
    	& \qquad  0,
    	 & & \mathrm{otherwise}.
    	\end{aligned} \right. 
    \end{align}
      \end{small}
\end{definition} 
Now, let $S_{\vect{u}}(\psi)$ be the power spectral density (PSD) of the sequence $\vect{u}$. According to the Wiener-Khinchin theorem, the PSD and ACF are Fourier transform
pairs \cite{Wiener1930}, i.e., 
\begin{equation}
   S_{\vect{u}}(\psi) = \sum_{\xi = -N+1}^{N-1}R_{\vect{u}}[\xi]\, e^{-j2\pi \psi \xi}.
\end{equation}For a Golay complementary pair $(\vect{u},\tilde{\vect{u}})$, we have 
\begin{equation}
\begin{aligned}
   S_{\vect{u}}(\psi) + S_{\tilde{\vect{u}}}(\psi) &= \sum_{\xi = -N+1}^{N-1} \left( R_{\vect{u}}[\xi] + R_{\tilde{\vect{u}}}[\xi]\right)\, e^{-j2\pi \psi \xi} \\
    & =2N \sum_{\xi = -N+1}^{N-1} \delta[\xi]\, e^{-j2\pi \psi \xi} = 2N. 
   \end{aligned}
\end{equation}We can see that the power spectra of a Golay complementary sequence pair add up to a constant.

We can define a Golay complementary array pair \cite{jedwab2007golay} in the following similar way.

\begin{definition}[Golay complementary array pair]
\label{def:Golay-array_pair}
    Unimodular arrays $\vect{U} \in \mathbb{C}^{N_1\times N_2}$ and $\tilde{\vect{U}} \in \mathbb{C}^{N_1 \times N_2}$ form a Golay complementary array pair if and only if 
    \begin{align}
    \label{eqn:conditionGolay2}
    	R_{\vect{U}}[\xi_1,\xi_2] + R_{\tilde{\vect{U}}}[\xi_1,\xi_2] &= 2N_1 N_2  \delta [\xi_1] \delta[\xi_2],
    \end{align}with $R_{\vect{U}} [\xi_1,\xi_2]$ being the ACF of $\vect{U}$ given in \eqref{eq:ACF2} at the top of the next page.
      \begin{figure*}[t]
    \begin{small}
    \begin{align}
    \label{eq:ACF2}
    	R_{\vect{U}}[\xi_1,\xi_2] = \left\{ 
    	\begin{aligned} 
    	&\sum\limits_{n_1=1}^{N_1-\xi_1} \sum\limits_{n_2=1}^{N_2-\xi_2} [\vect{U}]_{n_1,n_2} [\vect{U}]_{n_1+\xi_1,n_2+\xi_2}^*, 
    	 & & \xi_1=0,\ldots,N_1-1, \xi_2 = 0,\ldots,N_2-1,\\
    	&\sum\limits_{n_1=1}^{N_1+\xi_1} \sum\limits_{n_2=1}^{N_2-\xi_2} [\vect{U}]_{n_1-\xi_1,n_2} [\vect{U}]_{n_1,n_2+\xi_2}^*, 
    	 & & \xi_1=-N_1+1,\ldots,-1, \xi_2 = 0,\ldots,N_2-1,\\
      &\sum\limits_{n_1=1}^{N_1-\xi_1} \sum\limits_{n_2=1}^{N_2+\xi_2} [\vect{U}]_{n_1,n_2-\xi_2} [\vect{U}]_{n_1+\xi_1,n_2}^*, 
    	 & & \xi_1=0,\ldots,N_1-1, \xi_2 = -N_2+1,\ldots,-1,\\
      &\sum\limits_{n_1=1}^{N_1+\xi_1} \sum\limits_{n_2=1}^{N_2+\xi_2} [\vect{U}]_{n_1-\xi_1,n_2-\xi_2} [\vect{U}]_{n_1,n_2}^*, 
    	 & & \xi_1=-N_1+1,\ldots,-1, \xi_2 = -N_2+1,\ldots,-1,\\
    	& \qquad \qquad \qquad 0,
    	 & & \mathrm{otherwise}.
    	\end{aligned} \right. 
    \end{align}
      \end{small}
        \hrulefill
      \end{figure*}

\end{definition} 

Taking similar steps as before, we can show that the sum of the PSDs of a Golay complementary array pair is a constant:
\begin{equation}
\label{eq:PSD_2D}
    S_{\vect{U}}(\psi_1,\psi_2) + S_{\tilde{\vect{U}}}(\psi_1,\psi_2) = 2N_1 N_2.
\end{equation}

\subsection{RIS Configuration Design}
The following proposition shows the connection between Golay complementary array pairs and dual-polarized RIS phase configurations. 

\begin{proposition}
\label{prop:Golay_pair}
Consider a dual-polarized RIS with the phase configuration matrices $\boldsymbol{\Upsilon}_{\mathrm{p}} \in \mathbb{C}^{N_y \times \Tilde{N}_z},\, \mathrm{p} \in \{\mathrm{H},\mathrm{V}\}$, where the first column of $\boldsymbol{\Upsilon}_{\mathrm{p}}$ is formed by the first $N_\mathrm{y}$ entries of $\boldsymbol{\phi}_{\mathrm{p}}$, the second column consists of the second $N_\mathrm{y}$ entries of $\boldsymbol{\phi}_{\mathrm{p}}$ and so on. The RIS radiates a uniformly broad beam if and only if $(\boldsymbol{\Upsilon}_{\mathrm{H}},\boldsymbol{\Upsilon}_{\mathrm{V}})$ form a Golay complementary array pair.

\end{proposition}

\begin{IEEEproof}
  To radiate a uniformly broad beam, the per-polarization configurations of the RIS must satisfy \eqref{eq:PDAF_constant}. Using \eqref{eq:PDAF}, we can rewrite the condition in \eqref{eq:PDAF_constant} as  

\begin{align}
\label{eq:PDAF_constant_expanded}
  &\left| \sum_{n_z=1}^{\tilde{N}_z} \sum_{n_y=1}^{N_y}  [\boldsymbol{\Upsilon}_{\mathrm{H}}]_{n_y,n_z} e^{-j \left((n_y - 1)\hat{\psi}_y + (2n_z - 1)\hat{\psi}_z \right)}\right|^2 + \nonumber \\&\left| \sum_{n_z=1}^{\tilde{N}_z} \sum_{n_y=1}^{N_y}  [\boldsymbol{\Upsilon}_{\mathrm{V}}]_{n_y,n_z} e^{-j \left((n_y - 1)\hat{\psi}_y + 2(n_z - 1)\hat{\psi}_z \right)}\right|^2
  = c
 \end{align}
in which  
\begin{equation}
   \hat{\psi}_y = \psi_y + \tilde{\psi}_y,~~\hat{\psi}_z = \psi_z + \Tilde{\psi}_z,
\end{equation}where $\Tilde{\psi}_y$ and $\Tilde{\psi}_z$ are 
\begin{equation}
     \tilde{\psi}_y = \frac{2\pi}{\lambda}\Delta_y \sin(\tilde{\varphi}) \cos(\tilde{\theta}),\quad~\tilde{\psi_z} = \frac{2\pi}{\lambda} \Delta_z \sin(\tilde{\theta}).
\end{equation}
 The terms inside the magnitudes in  \eqref{eq:PDAF_constant_expanded} are the 2D discrete-space Fourier transform of  $\boldsymbol{\Upsilon}_{\mathrm{H}}$ and $\boldsymbol{\Upsilon}_{\mathrm{V}}$. Since the magnitude squared of the Fourier transform of an array equals its PSD  \cite[Chapter 4]{Stein2000}, \cite[Chapter 4]{semmlow2011}, \eqref{eq:PDAF_constant_expanded} is equivalent to  
\begin{equation}
\label{eq:sum_spectral_density}
 S_{\boldsymbol{\Upsilon}_{\mathrm{H}}}(\hat{\psi}_y,\hat{\psi}_z) + S_{\boldsymbol{\Upsilon}_{\mathrm{V}}}(\hat{\psi}_y,\hat{\psi}_z) = c.
\end{equation} Taking inverse Fourier transform from both sides of \eqref{eq:sum_spectral_density}, we arrive at 
\begin{equation}
\label{eq:sum_ACF_RISconfigs}R_{\boldsymbol{\Upsilon}_{\mathrm{H}}}[\xi_1,\xi_2] + R_{\boldsymbol{\Upsilon}_{\mathrm{V}}}[\xi_1,\xi_2]  = c\delta [\xi_1]\delta[\xi_2], 
\end{equation}which signifies that the ACFs of matrices $(\boldsymbol{\Upsilon}_{\mathrm{H}},\boldsymbol{\Upsilon}_{\mathrm{V}})$ must add up to zero except at $\xi_1 = \xi_2 = 0$ if we want the RIS to radiate a uniformly broad beam (with a spatially flat array factor). Therefore, $(\boldsymbol{\Upsilon}_{\mathrm{H}},\boldsymbol{\Upsilon}_{\mathrm{V}})$ must form a Golay complementary array pair in which case we have $c = 2 \Tilde{N}_z N_y = N_z N_y$. Hence, if the dual-polarized RIS radiates a broad beam, then, its phase configuration matrices form a Golay complementary array pair. 

On the other hand, if the RIS phase configuration matrices $(\boldsymbol{\Upsilon}_{\mathrm{H}},\boldsymbol{\Upsilon}_{\mathrm{V}})$ constitute a Golay complementary array pair, \eqref{eq:PSD_2D} yields that 
\begin{equation}
\label{eq:sum_PSD}
 S_{\boldsymbol{\Upsilon}_{\mathrm{H}}}(\psi_1,\psi_2) + S_{\boldsymbol{\Upsilon}_{\mathrm{V}}}(\psi_1,\psi_2) = 2 \Tilde{N}_z N_y = N_z N_y.
\end{equation}Writing the PSDs in \eqref{eq:sum_PSD} as the magnitude squared of the Fourier transform of the matrices, we arrive at 
\begin{align}    \label{eq:PDAF_constant_expanded2}
 & \left| \sum_{n_z=1}^{\tilde{N}_z} \sum_{n_y=1}^{N_y}  [\boldsymbol{\Upsilon}_{\mathrm{H}}]_{n_y,n_z} e^{-j \left(n_y \psi_1 + n_z \psi_2 \right)}\right|^2 + \nonumber \\& \left| \sum_{n_z=1}^{\tilde{N}_z} \sum_{n_y=1}^{N_y}  [\boldsymbol{\Upsilon}_{\mathrm{V}}]_{n_y,n_z} e^{-j \left(n_y \psi_1 + n_z \psi_2\right)}\right|^2 =N_z N_y.
\end{align}We set $\psi_1 = \hat{\psi}_y $ and $\psi_2 = 2\hat{\psi}_z$. Then, we multiply the terms inside the magnitude in the first line of \eqref{eq:PDAF_constant_expanded2} by $e^{j(\hat{\psi}_y + \hat{\psi}_z)}$ and those inside the magnitude in the second line of \eqref{eq:PDAF_constant_expanded2} by $e^{j(\hat{\psi}_y + 2\hat{\psi}_z)}$. Thus, we end up at 
\begin{equation}
   A(\varphi,\theta) = N_z N_y, 
\end{equation}which indicates that an RIS whose configuration matrices form a Golay array pair produces a uniformly broad beam with a spatially flat power-domain array factor. 
\end{IEEEproof}

\begin{remark}
  Based on Weyl's identity, any spherical wave can be represented by the summation of a number of plane waves \cite{pizzo2020degrees,kosasih2024roles}. This implies that a broad beam over all far-field LoS angles will cover any potential near-field LoS user. 
\end{remark}

\begin{figure*}
	\centering
	\begin{subfigure}{0.33\textwidth}
		\centering
		\includegraphics[width=\columnwidth]{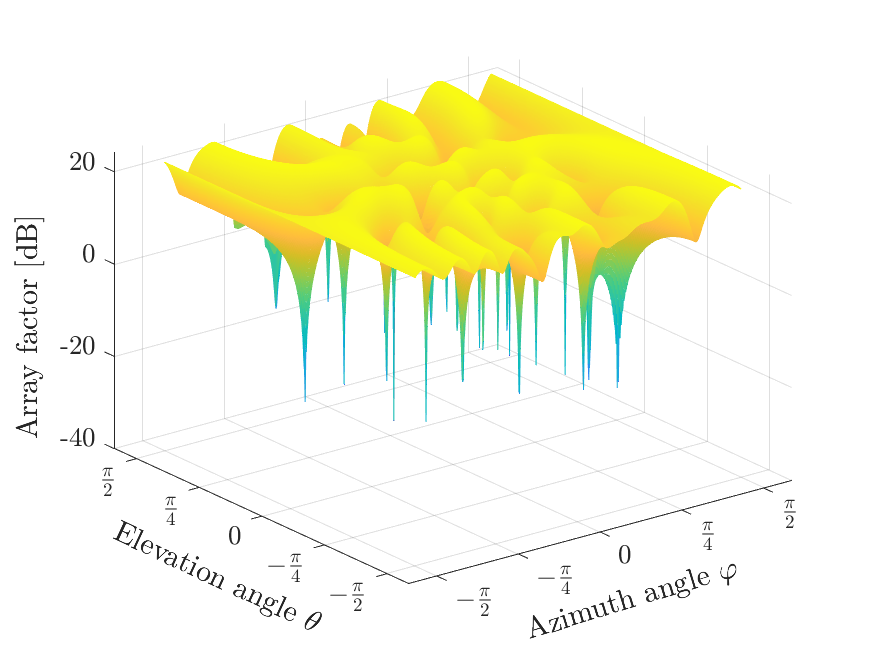}
		\caption{Polarization H}
		\label{fig:AF-polH}
	\end{subfigure}%
	\begin{subfigure}{0.33\textwidth}
		\centering
		\includegraphics[width=\columnwidth]{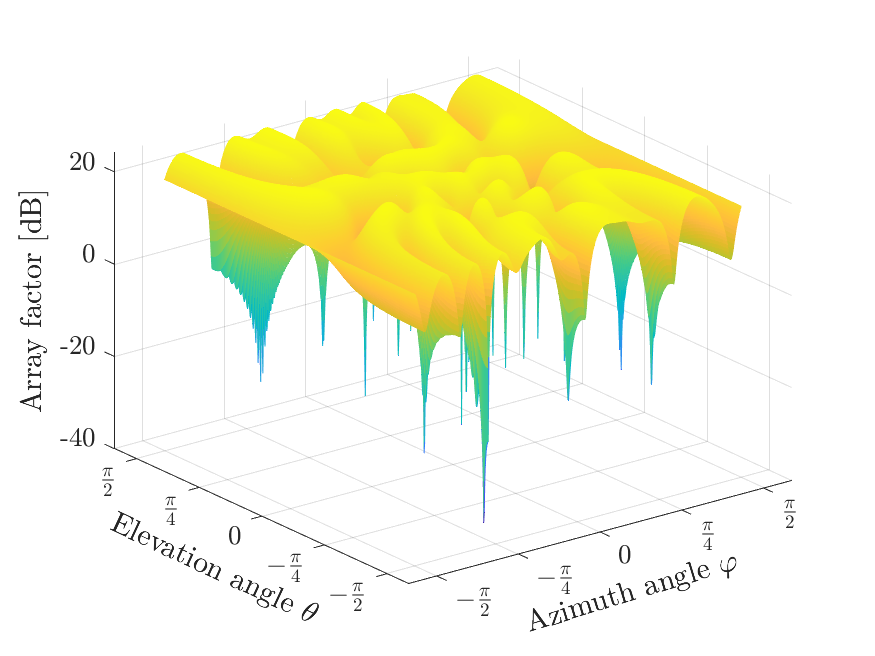}
		\caption{Polarization V}
		\label{fig:AF-polV}
	\end{subfigure}%
 \begin{subfigure}{0.33\textwidth}
		\centering
		\includegraphics[width=\columnwidth]{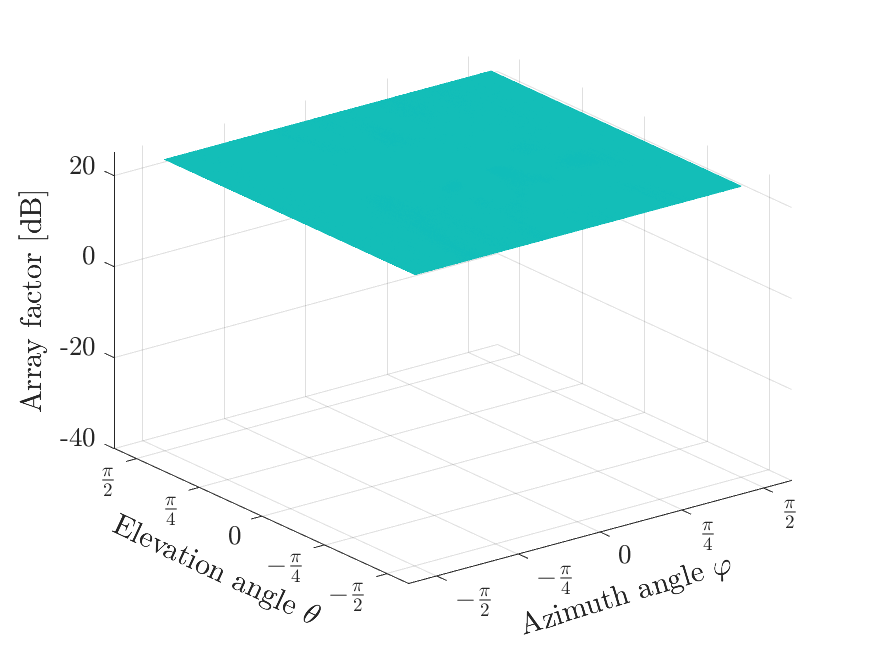}
		\caption{Total}
		\label{fig:AF-tot}
	\end{subfigure}
	\caption{Illustration of the per-polarization and total power-domain array factor for a dual-polarized RIS for which the phase configuration matrices of the two polarizations form a Golay complementary array pair. The total gain is flat, while the individual polarizations have peaks and valleys. The considered RIS has $256$ elements with $N_y = N_z = 16$ and an inter-element spacing of $\Delta_y = \Delta_z = \lambda/4$.}
 \label{fig:AF}
\end{figure*}

\begin{figure}
    \centering
    \includegraphics[width=0.9\columnwidth]{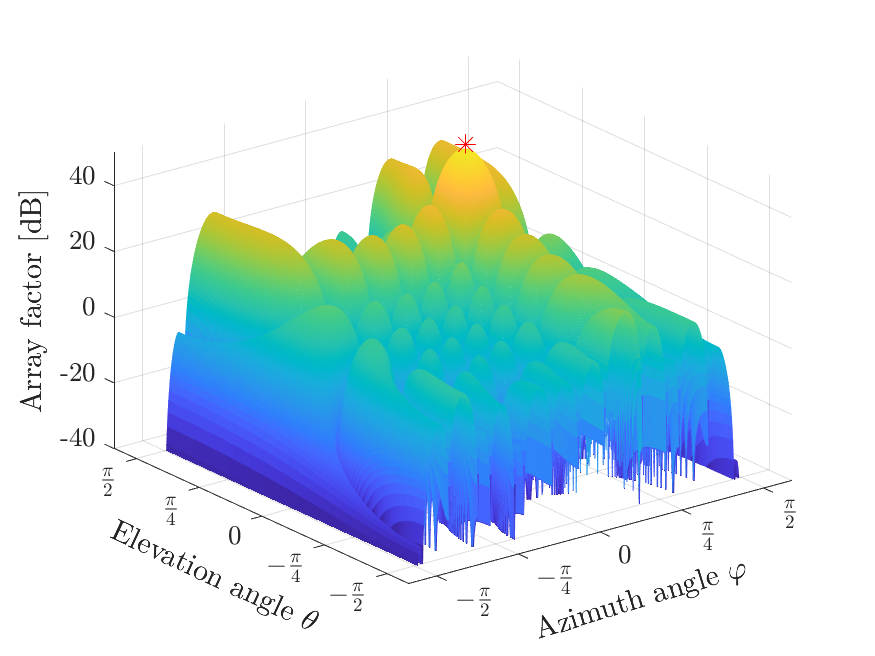}
    \caption{Total power-domain array factor with user-specific beamforming. The setup is the same as that in Fig.~\ref{fig:AF} and the RIS points the reflected beam towards $\varphi = \theta = \pi/6$. }
    \label{fig:AF_UE}
\end{figure}

\section{Construction of Golay Complementary Array Pairs for Broad Dual-Polarized Beamforming}
\label{sec:Golay_construction}
Many Golay complementary sequence pairs have been identified in previous literature via exhaustive numerical search, though it is known that for some sequence lengths, no Golay sequence pair exists \cite{Holzmann94}. Here, we describe how to construct Golay complementary array pairs based on known Golay complementary sequence pairs.    
\begin{proposition}
    \label{prop:Golay-expansion}
Let $(\vect{u}_1,\tilde{\vect{u}}_1)$ and $(\vect{u}_2,\tilde{\vect{u}}_2)$ be two pairs of Golay complementary sequences of lengths $L_1$ and $L_2$, respectively. Then, $(\vect{U},\tilde{\vect{U}})$ formed by either
\begin{equation}
\begin{aligned}
    	\label{eq:Golay-expansion1}
    	&\vect{U} = \begin{bmatrix}
    		\vect{u}_1 \vect{u}_2^{\Ttran}\\
    		-\tilde{\vect{u}}_1\tilde{\vect{u}}_2^{\Htran}\vect{E}_{L_2}
    	\end{bmatrix} \in \mathbb{C}^{2L_1 \times L_2},\\
    &\tilde{\vect{U}} = \begin{bmatrix}
    \vect{u}_1\tilde{\vect{u}}_2^{\Ttran}\\
    		\tilde{\vect{u}}_1 \vect{u}_2^{\Htran}\vect{E}_{L_2}
    	\end{bmatrix}\in \mathbb{C}^{2L_1 \times L_2},
    \end{aligned}
    \end{equation}or
    \begin{equation}
        \begin{aligned}
        \label{eq:Golay-expansion2}
       &\vect{U} = \left[\vect{u}_1 \vect{u}_2^{\Ttran},\; -\tilde{\vect{u}}_1\tilde{\vect{u}}_2^{\Htran}\vect{E}_{L_2}\right] \in \mathbb{C}^{L_1 \times 2L_2}, \\
       & \tilde{\vect{U}} = \left[\vect{u}_1\tilde{\vect{u}}_2^{\Ttran},\; \tilde{\vect{u}}_1 \vect{u}_2^{\Htran}\vect{E}_{L_2}\right]\in \mathbb{C}^{L_1 \times 2L_2} 
        \end{aligned}
    \end{equation}
is a Golay complementary array pair, where $\vect{E}_L$ is an $L \times L$ matrix having ones on the anti-diagonal and zeros elsewhere.  
    	
\end{proposition}
\begin{IEEEproof}
    The proof is similar to the proof in \cite[Theorem III.3]{Li2021Golay}. In summary, to prove that the array pair in \eqref{eq:Golay-expansion1} is a Golay complementary pair, we re-write $\vect{U}$ and $\tilde{\vect{U}}$ as 
    \begin{equation}
        \begin{aligned}
        &\vect{U} = \begin{bmatrix} \vect{u}_1 \\ \vect{0}_{L_1 \times 1} 
        \end{bmatrix} \vect{u}_2^{\Ttran} - \begin{bmatrix}
            \vect{0}_{L_1\times 1} \\
            \tilde{\vect{u}}_1
   \end{bmatrix}\tilde{\vect{u}}_2^{\Htran}\vect{E}_{L_2}  \\
       &\tilde{\vect{U}} = \begin{bmatrix} \vect{u}_1 \\ \vect{0}_{L_1 \times 1} 
        \end{bmatrix} \tilde{\vect{u}}_2^{\Ttran} + \begin{bmatrix}
            \vect{0}_{L_1\times 1} \\
            \tilde{\vect{u}}_1
   \end{bmatrix}\vect{u}_2^{\Htran}\vect{E}_{L_2}, 
        \end{aligned} 
    \end{equation}and show that 
    \begin{equation}
    \label{eq:Golay_pair_condition}
        R_{\vect{U}}[\xi_1,\xi_2] + R_{\tilde{\vect{U}}}[\xi_1,\xi_2] = 4L_1 L_2 \delta[\xi_1]\delta[\xi_2]
    \end{equation}using the properties of cross-correlation and auto-correlation operations given in \cite[Lemma II.1, Lemma III.1, Lemma III.2]{Li2021Golay}. Similarly, the array pair in \eqref{eq:Golay-expansion2} can be proved to be a Golay complementary pair by re-writing $\vect{U}$ and $\tilde{\vect{U}}$ as
    \begin{equation}
     \begin{aligned}
        &\vect{U} = \vect{u}_1 \left[\vect{u}_2^{\Ttran}, \; \vect{0}_{1\times L_2} \right] - \tilde{\vect{u}}_1 \left[\vect{0}_{1 \times L_2},\; \tilde{\vect{u}}_2^{\Htran}\vect{E}_{L_2} \right]  \\
        &\tilde{\vect{U}} = \vect{u}_1 \left[\tilde{\vect{u}}_2^{\Ttran}, \; \vect{0}_{1\times L_2} \right] +  \tilde{\vect{u}}_1 \left[\vect{0}_{1 \times L_2},\; \vect{u}_2^{\Htran}\vect{E}_{L_2} \right]
     \end{aligned}   
    \end{equation}and showing that \eqref{eq:Golay_pair_condition} holds. 
\end{IEEEproof} 

We now provide an illustrative example to demonstrate the spatially flat power-domain array factor produced by a dual-polarized RIS that is configured by a Golay complementary array pair.  We consider an RIS with $N_y = N_z = 16$ and $\Delta_y = \Delta_z = \lambda/4$. The AoAs to the RIS are set as $\Tilde{\varphi} = \pi/3$ and $\Tilde{\theta} = \pi/3$. Each polarization of the RIS consists of $128$ elements in the form of a $16 \times 8$ UPA. We use Proposition~\ref{prop:Golay-expansion} to obtain the phase configuration matrices of the RIS using two known Golay complementary sequence pairs of length $8$. Specifically, we utilize the following two pairs of Golay sequences \cite{Holzmann94} for the construction of a Golay complementary array pair:
\begin{equation}
    \begin{aligned}
     &\vect{u}_1 = \exp \left(j\left[0,0,0,0,0,0,\pi,\pi,0\right]^{\Ttran}\right),\\
     &\tilde{\vect{u}}_1 = \exp \left(j\left[0,0,\pi,\pi,0,\pi,0,\pi\right]^{\Ttran}\right),
    \end{aligned}
\end{equation}and
\begin{equation}
    \begin{aligned}
     &\vect{u}_2 = \exp \left(j\left[0,0,0,0,\frac{\pi}{2},\frac{-\pi}{2},\frac{-\pi}{2},\frac{\pi}{2}\right]^{\Ttran}\right),\\
     &\tilde{\vect{u}}_2 = \exp \left(j\left[0,0,\pi,\pi,\frac{\pi}{2},\frac{-\pi}{2},\frac{\pi}{2},\frac{-\pi}{2}\right]^{\Ttran}\right).
    \end{aligned}
\end{equation}
Then, the two phase configuration matrices of the RIS for polarizations $\mathrm{H}$ and $\mathrm{V}$ are formed by \eqref{eq:Golay-expansion1}. Fig.~\ref{fig:AF} illustrates the power-domain array factor for the considered RIS, where the array factors for polarizations $\mathrm{H}$ and $\mathrm{V}$ are given in Fig.~\ref{fig:AF-polH} and Fig.~\ref{fig:AF-polV}, respectively. 
Each of these array factors has large values at many angles; however, there are also many valleys in between the peaks, which is unavoidable for uni-polarized transmission.
Fig.~\ref{fig:AF-tot} shows the total power-domain array factor obtained by summing up the power-domain array factors of the two polarizations. It can be observed that the total power-domain array factor is spatially flat over $\varphi \in [-\pi/2, \pi/2]$ and $\theta \in [-\pi/2, \pi/2]$, verifying the theoretical result given in Proposition~\ref{prop:Golay_pair}.
The constant total power-domain array factor is  $A(\varphi,\theta) = 10\log(2N) = 24.08\,$dB for all the considered azimuth and elevation angles. 

We now compare the total power-domain array factor of the broad beamforming in Fig.~\ref{fig:AF-tot} with that of user-specific beamforming when the reflected beam from the RIS is focused towards one user which is assumed to be located in the angular direction $\varphi =\theta = \pi/6$.  Fig.~\ref{fig:AF_UE} depicts the total power-domain array factor for the latter scenario. The maximum array factor, marked with a red star, is obtained at $\varphi =\theta = \pi/6$, and equals $10\log(2N^2) = 45.15\,$dB. Although this value is almost twice that achievable with broad beamforming,
 we can observe notable fluctuations of the array factor value over different angles and the received power in some directions is negligible. Specifically, the average array factor value over all directions is $3.61$\,dB which is much smaller than the array factor we get with broad beamforming.

In addition to utilizing Golay complementary sequence pairs for constructing Golay complementary array pairs, we can also use Golay complementary array pairs of smaller sizes to construct larger pairs. The following proposition elaborates on this. 
\begin{proposition}
  Let $(\vect{U}_1,\tilde{\vect{U}}_1)$ be a Golay array pair of size $L_1 \times J_1$ and $(\vect{U}_2,\tilde{\vect{U}}_2)$ represent a Golay array pair of size $L_2 \times J_2$. An expanded Golay array pair can be constructed as either 
  \begin{equation}
      \begin{aligned}
        &\vect{W} = \begin{bmatrix}
           \vect{U}_1 \otimes \vect{U}_2 \\
           - \tilde{\vect{U}}_1 \otimes \vect{E}_{L_2}\tilde{\vect{U}}_2^* \vect{E}_{J_2}
        \end{bmatrix} \in \mathbb{C}^{2L_1 L_2 \times J_1 J_2}  \\
        &\Tilde{\vect{W}} = \begin{bmatrix}
           \vect{U}_1 \otimes \tilde{\vect{U}}_2 \\
            \tilde{\vect{U}}_1 \otimes \vect{E}_{L_2}\vect{U}_2^* \vect{E}_{J_2}
        \end{bmatrix} \in \mathbb{C}^{2L_1 L_2 \times J_1 J_2}
      \end{aligned}
  \end{equation}
  or
  \begin{equation}
      \begin{aligned}
       &\vect{W} = \left[\vect{U}_1 \otimes \vect{U}_2, \;  - \tilde{\vect{U}}_1 \otimes \vect{E}_{L_2}\tilde{\vect{U}}_2^* \vect{E}_{J_2}\right] \in \mathbb{C}^{L_1 L_2 \times 2J_1 J_2}  \\
       & \Tilde{\vect{W}} = \left[\vect{U}_1 \otimes \tilde{\vect{U}}_2,\;  \tilde{\vect{U}}_1 \otimes \vect{E}_{L_2}\vect{U}_2^* \vect{E}_{J_2}\right] \in \mathbb{C}^{L_1 L_2 \times 2J_1 J_2}.
      \end{aligned}
  \end{equation}
\end{proposition}

\begin{IEEEproof}
  The proof is similar to the proof in  \cite[Theorem IV.2]{Du2022constructions}.  
\end{IEEEproof}

\section{Broad-Beam Design via \texorpdfstring{$\epsilon$}~-Complementary Pairs }
\label{sec:epsilon_comp}
So far, we have designed RIS configuration pairs to have a uniformly broad reflected beam from the RIS, assuming a purely LoS channel between the BS and the RIS. Herein, we consider a general scenario where the BS-RIS channel have NLoS propagation paths as well. In this case, due to the varying amplitude of the channel between the BS and the RIS, the achievement of a uniformly broad radiated beam from the RIS cannot be guaranteed. Hence, we propose a new method for producing a practically broad reflected beam from the RIS using stochastic optimization. 

\begin{remark}
  In a purely LoS scenario, there is no leakage between the polarizations and a signal transmitted in polarization $\mathrm{p}$ reaches the RIS with no polarization change. In such a case, the received signal model is similar to that of two traditional uni-polarized RIS-assisted systems that are operated in parallel, as is clear from \eqref{eq:received_signal}. However, when considering an arbitrary BS-RIS channel with NLoS components, the cross-polarization leakage is generally non-zero such that a portion of the signal transmitted in polarization $\mathrm{H}$ reaches the RIS with polarization $\mathrm{V}$ and vice versa. 
\end{remark}

 In a general dual-polarized RIS-assisted system, the channel between the BS and the RIS is given by \cite{Chen2021}
\begin{equation}
 \vect{H} = \begin{bmatrix}
    		\hat{\vect{H}}_{1,1}& \ldots &\hat{\vect{H}}_{1,M} \\
      \vdots & \ddots & \vdots
      \\
      \hat{\vect{H}}_{N,1}& \ldots &\hat{\vect{H}}_{N,M} 
    	\end{bmatrix} \in \mathbb{C}^{2N \times 2M},
\end{equation}where $\hat{\vect{H}}_{n,m} \in \mathbb{C}^{2\times 2}$ is the channel between the  $n$th pair of dual-polarized RIS elements and the $m$th pair of dual-polarized BS antennas, given by

\begin{align}
\hat{\vect{H}}_{n,m} &=\left(\begin{bmatrix} 
\sqrt{1-q} & \sqrt{q} \\
\sqrt{q} & \sqrt{1-q}
\end{bmatrix}  \odot \begin{bmatrix}
    h_{n,m}^{\mathrm{H},\mathrm{H}} & h_{n,m}^{\mathrm{H},\mathrm{V}} \\
    h_{n,m}^{\mathrm{V},\mathrm{H}} & h_{n,m}^{\mathrm{V},\mathrm{V}}
\end{bmatrix}\right)
\end{align}
where $q = \frac{1}{1+\mathcal{X}}$ and $\mathcal{X}$ is the cross-polarization discrimination (XPD), which is the ratio of the power received in the desired polarization to the power received in the undesired polarization.  
Moreover, $h_{n,m}^{\mathrm{p},\mathrm{q}}$ represents the channel between the $n$th $\mathrm{p}$-polarized RIS element and $m$th $\mathrm{q}$-polarized BS antenna for $\mathrm{p}, \mathrm{q} \in \{\mathrm{H},\mathrm{V}\}$.
The received signal at a potential user located in the azimuth angle $\varphi$ and elevation angle $\theta$ is obtained as (40) at top of the next page. $\Tilde{\vect{H}}_{\mathrm{p},\mathrm{q}} \in \mathbb{C}^{N \times M}$ indicates the channel between the $\mathrm{p}$-polarized RIS elements and $\mathrm{q}$-polarized BS antennas, with $h^{\mathrm{p},\mathrm{q}}_{n,m}$ as its entry in the $n$th row and $m$th column. 

For the purpose of exposition and drawing the fundamental design insights, we assume perfect isolation between the polarizations herein, i.e., $q = 0$. So, the received signal in polarization $\mathrm{p}$ will be given by \eqref{eq:received_signal}, where $ \vect{H}_{\mathrm{H}} = \Tilde{\vect{H}}_{\mathrm{H,H}}$ and $ \vect{H}_{\mathrm{V}} = \Tilde{\vect{H}}_{\mathrm{V,V}}$. All the provided analyses are readily applicable to the case with arbitrary $q$.
\begin{figure*}
\begin{subequations}
    \begin{align}
    r_{\mathrm{H}} &= \sqrt{P_{\mathrm{T}} \Tilde{\beta}\beta G_{\mathrm{R},0}(\varphi,\theta)} \vect{a}_{\mathrm{H}}^{\Ttran}\boldsymbol{\Phi}_{\mathrm{H}} \left(\sqrt{1-q} \Tilde{\vect{H}}_{\mathrm{H,H}} \vect{w}_{\mathrm{H}}+ \sqrt{q}\tilde{\vect{H}}_{\mathrm{H,V}}\vect{w}_{\mathrm{V}}\right) + n_{\mathrm{H}}  \\
     r_{\mathrm{V}} & = \sqrt{P_{\mathrm{T}} \Tilde{\beta}\beta G_{\mathrm{R},0}(\varphi,\theta)}\vect{a}_{\mathrm{V}}^{\Ttran}\boldsymbol{\Phi}_{\mathrm{V}} \left(\sqrt{1-q} \Tilde{\vect{H}}_{\mathrm{V,V}} \vect{w}_{\mathrm{V}}+ \sqrt{q}\tilde{\vect{H}}_{\mathrm{V,H}}\vect{w}_{\mathrm{H}}\right) + n_{\mathrm{V}}
    \end{align}
    \end{subequations}
\hrulefill
      \end{figure*}

To maximize the received power at the RIS, the beamforming vectors at the BS are set as $\vect{w}_{\mathrm{p}} = \vect{f}_{\mathrm{p}}$, where $\vect{f}_{\mathrm{p}}$ is the eigenvector corresponding to the largest eigenvalue of $\vect{H}^{\Htran}_{\mathrm{p}} \vect{H}_{\mathrm{p}}$. Setting $\vect{h}_{\mathrm{p}} = \vect{H}_{\mathrm{p}}\vect{f}_{\mathrm{p}}$, the received signal in polarization $\mathrm{p}$ is given by  \begin{equation}
\label{eq:received_signal_arbitrary}
    r_{\mathrm{p}} = \sqrt{P_{\mathrm{T}} \Tilde{\beta} \beta G_{\mathrm{R},0}(\varphi,\theta)} \left( \vect{h}_{\mathrm{p}} \odot \boldsymbol{\phi}_{\mathrm{p}} \right)^{\Ttran} \vect{a}_{\mathrm{p}}(\varphi,\theta) + n_{\mathrm{p}},
\end{equation}
and SNR at an arbitrary user located at $(\varphi,\theta)$ is computed as 
\begin{equation}
\label{eq:SNR_arbitrary}
     \mathrm{SNR} = \gamma(\varphi,\theta) \sum_{\mathrm{p} = {\mathrm{H,\mathrm{V}}}}\left|\left(\vect{h}_{\mathrm{p}} \odot \boldsymbol{\phi}_{\mathrm{p}}\right)^{\Ttran} \vect{a}_{\mathrm{p}}(\varphi,\theta)\right|^2
\end{equation} with 
\begin{equation}
   \gamma(\varphi,\theta) = \frac{P_{\mathrm{T}} 
 \Tilde{\beta} \beta G_{\mathrm{R},0}(\varphi,\theta)}{\sigma^2}.
\end{equation}
We can then define the new power-domain array factor as
\begin{equation}
\label{eq:array_factor_arbitrary}
    A(\varphi,\theta) = \left|\hat{\boldsymbol{\phi}}_{\mathrm{H}}^{\Ttran} \vect{a}_{\mathrm{H}}(\varphi,\theta)\right|^2 + \left|\hat{\boldsymbol{\phi}}_{\mathrm{V}}^{\Ttran} \vect{a}_{\mathrm{V}}(\varphi,\theta)\right|^2,
\end{equation}where $\hat{\boldsymbol{\phi}}_{\mathrm{p}} = \vect{h}_{\mathrm{p}} \odot \boldsymbol{\phi}_{\mathrm{p}}$.
\begin{remark}
  The extension of the considered model to wideband scenarios is straightforward. Specifically, in a wideband system with multiple subcarriers, $\vect{h}_{\mathrm{p}}$ in \eqref{eq:received_signal_arbitrary} can be replaced by $\sum_{i=1}^{N_{\mathrm{sub}}}\vect{h}_{\mathrm{p},i}$, where $N_{\mathrm{sub}}$ is the number of subcarriers and $\vect{h}_{\mathrm{p},i} = \vect{H}_{\mathrm{p},i}\vect{f}_{\mathrm{p},i}$ with $\vect{H}_{\mathrm{p},i}$ and $\vect{f}_{\mathrm{p},i}$ being the channel between the BS and the RIS on the $i$-th subcarrier and the eigenvector corresponding to the largest eigenvalue of $\vect{H}_{\mathrm{p},i}^{\Htran} \vect{H}_{\mathrm{p},i}$, respectively. However, an important consideration is that RIS elements act as linear-phase filters only over a specific bandwidth interval. Outside this interval, RIS acts non-linearly, leading to beam squint effects. Therefore, it is important to take the RIS linearity interval into account when studying wideband systems. 
\end{remark}

Similar to the case with LoS BS-RIS channel, considered in the previous sections, the objective is to design the RIS phase configurations such that the power-domain array factor  becomes spatially flat over the possible observation angles. To simplify the presentation, we first consider a horizontal ULA-type RIS and then extend our analyses to the scenario with a UPA-type RIS. 
For a ULA-type dual-polarized RIS in which adjacent elements have opposite polarizations, the array response vectors are given by \cite{ParisaBroadBeamLetter}

\begin{align}
 \label{eq:array_response_H_ULA} \vect{a}_{\mathrm{H}}(\varphi) &= \left[1,e^{-j\frac{2\pi}{\lambda}2\Delta_y \sin(\varphi)},\ldots,e^{-j\frac{2\pi}{\lambda}2\Delta_y (N-1) \sin(\varphi)}  \right]^{\Ttran},  \\
  \vect{a}_{\mathrm{V}}(\varphi) &= e^{-j\frac{2\pi}{\lambda}\Delta_y \sin(\varphi)} \vect{a}_{\mathrm{H}}(\varphi), \label{eq:array_response_V_ULA}
\end{align}and the requirement for a spatially flat power-domain array factor is expressed as 
\begin{equation}
\label{eq:spatially_flat_ULA}
    A(\varphi) = \left|\hat{\boldsymbol{\phi}}_{\mathrm{H}}^{\Ttran} \vect{a}_{\mathrm{H}}(\varphi)\right|^2 + \left|\hat{\boldsymbol{\phi}}_{\mathrm{V}}^{\Ttran} \vect{a}_{\mathrm{V}}(\varphi)\right|^2 = c.
\end{equation} Since $\hat{\boldsymbol{\phi}}_{\mathrm{p}}^{\Ttran} \vect{a}_{\mathrm{p}}(\varphi)$ is the discrete-space Fourier transform of $\hat{\boldsymbol{\phi}}_{\mathrm{p}}$, following similar steps as in Section~\ref{sec:Golay_pairs}, the broad-beam condition in \eqref{eq:spatially_flat_ULA} can be expressed in terms of the ACFs of $\hat{\boldsymbol{\phi}}_{\mathrm{H}}$ and $\hat{\boldsymbol{\phi}}_{\mathrm{V}}$ as 
\begin{equation}
\label{eq:ACF_constant_arbitrary}
    R_{\hat{\boldsymbol{\phi}}_{\mathrm{H}}}[\xi] + R_{\hat{\boldsymbol{\phi}}_{\mathrm{V}}}[\xi] = c\delta [\xi].
\end{equation} We need to design the per-polarization RIS phase-shifts such that \eqref{eq:ACF_constant_arbitrary} is satisfied. However, as the NLoS channel components of the BS-RIS channel are involved in $\hat{\boldsymbol{\phi}}_{\mathrm{H}}$ and $\hat{\boldsymbol{\phi}}_{\mathrm{V}}$, it is not straightforward to find RIS configurations that strictly satisfy \eqref{eq:ACF_constant_arbitrary}. We thus introduce some relaxations to \eqref{eq:ACF_constant_arbitrary} and propose a stochastic optimization approach for finding a suitable RIS configuration pair. 
To this end, we relax the requirement of having a uniformly broad beam and allow for some level of imperfection in the sum of ACFs in \eqref{eq:ACF_constant_arbitrary}. Hence, we look for a pair of $(\hat{\boldsymbol{\phi}}_{\mathrm{H}}, \hat{\boldsymbol{\phi}}_{\mathrm{V}})$ that fulfills 
\begin{equation}
\label{eq:epsilon_complementary_condition}
    \left| R_{\hat{\boldsymbol{\phi}}_{\mathrm{H}}}[\xi] + R_{\hat{\boldsymbol{\phi}}_{\mathrm{V}}}[\xi] \right| \leq \epsilon,~~\forall \xi \neq 0,
\end{equation}
where $\epsilon>0$ is a small number.
This relaxation implies that we allow for a certain level of deviation from a uniformly broad beam. However, the deviation is guaranteed to be limited and controlled by the selection of $\epsilon$, making the reflected beam from the RIS practically indistinguishable from a uniformly broad beam. Here, we propose a heuristic algorithm for finding the RIS phase configurations so as to obtain a practically broad beam. 

We define the RIS configuration vector $\boldsymbol{\phi} = \left[\boldsymbol{\phi}_{\mathrm{H}}^{\Ttran},\boldsymbol{\phi}_{\mathrm{V}}^{\Ttran}\right]^{\Ttran}$ and consider the utility function 
\begin{equation}
    F(\boldsymbol{\phi}) = - \underset{\xi \in \mathcal{N}}{\max}\, \left| R_{\hat{\boldsymbol{\phi}}_{\mathrm{H}}}[\xi] + R_{\hat{\boldsymbol{\phi}}_{\mathrm{V}}}[\xi] \right|, 
\end{equation}
where $\mathcal{N} = \{-N+1,\ldots,N+1\} \setminus \{0\}$. The objective is to maximize this utility function which in turn minimizes the level of sidepeaks in the total ACF. The corresponding problem can be formulated as
\begin{align}
\mathbb{P:}~~\underset{\boldsymbol{\phi}}{\mathrm{maximize}}~~
  &F(\boldsymbol{\phi}) \notag \\
   \mathrm{subject \, to}~~&\omega_{\mathrm{p},n} \in [0,2\pi),~  \mathrm{p} \in \{\mathrm{H},\mathrm{V}\},\, n \in \{1,\ldots,N\}. \notag
\end{align}We define $\boldsymbol{\Omega} = \arg(\boldsymbol{\phi}) = [\omega_{\mathrm{H},1},\ldots, \omega_{\mathrm{H},N},\omega_{\mathrm{V},1},\ldots, \omega_{\mathrm{V},N}]^{\Ttran}$ for later use and let $\Omega_n$ be the $n$th entry of $\boldsymbol{\Omega}$ for $n = 1,\ldots,2N$. Our approach for solving Problem $\mathbb{P}$ is to gradually reduce the level of sidepeaks in the total ACF by sequentially refining the RIS phase shifts in an iterative manner until the sidepeak level falls within the tolerance threshold specified by $\epsilon$ or a maximum number of iterations is reached. We call the set of RIS phase shift configurations obtained in this way an \textit{$\epsilon$-complementary configuration pair} and the steps for finding it are provided in Algorithm \ref{alg:epsilon-comp-pairs}. 

\begin{algorithm}

\caption{Proposed algorithm for finding a pair of $\epsilon$-complementary phase configurations.}
\label{alg:epsilon-comp-pairs}
	\begin{algorithmic}[1]
 \STATEx{\textbf{Inputs:} $\vect{h}_{\mathrm{H}},~\vect{h}_{\mathrm{V}}$.}
		
		\STATE {Set the convergence threshold $\epsilon$, scaling factor $\alpha$, maximum number of iterations $L_1$ and $L_2$.}
		\STATE {Initialize $\boldsymbol{\Omega} \sim \mathcal{U}[0, 2\pi)^{2N}$.}
		\STATE {Set $l_1=0$.}
		    		
        \WHILE{$|F(\boldsymbol{\phi})| > \epsilon ~\&~ l_1 < L_1$ }
        \STATE $l_1 \leftarrow l_1+1$.
        \STATE Create a phase increment vector $\Delta \boldsymbol{\Omega} \sim \mathcal{U}[0, 2\pi)^{2N}$.
        \FOR{ $n = 1:2N $}
        \STATE{Set $\mu \leftarrow F(\boldsymbol{\phi})$.}
        \STATE{Set $l_2 = 0$.}
        \WHILE{$F(\boldsymbol{\phi}) \leq \mu ~\&~ l_2 < L_2$}
        \STATE{$l_2 \leftarrow l_2 +1$.}
        \STATE{$\Omega_n \leftarrow \Omega_n + \Delta \Omega_n$.}
        \IF{$F(\boldsymbol{\phi}) < \mu$}
        \STATE{$\Omega_n \leftarrow \Omega_n - 2\Delta \Omega_n$.}
        \IF{$F(\boldsymbol{\phi}) < \mu$}
        \STATE{$\Omega_n \leftarrow \Omega_n + \Delta \Omega_n,~\Delta \Omega_n \leftarrow \alpha \Delta \Omega_n $.}
        \ELSE
        \STATE{Accept new phase shift.}
        \ENDIF
        \ELSE
        \STATE{Accept new phase shift.}
        \ENDIF
        \ENDWHILE

        \ENDFOR
		\ENDWHILE
		\STATEx{\textbf{Output:} $\boldsymbol{\phi} = \left[\boldsymbol{\phi}_{\mathrm{H}}^{\Ttran},\boldsymbol{\phi}_{\mathrm{V}}^{\Ttran}  \right]^{\Ttran}$.}
	\end{algorithmic}
\end{algorithm} 

This is how Algorithm~\ref{alg:epsilon-comp-pairs} works. We first initialize all the RIS phase shifts, compute the utility function for the initialized phase shifts and enter a \textit{while} loop in which the phase shifts are iteratively refined until the utility function value becomes less than the threshold specified by $\epsilon$ or the maximum number of iterations is reached.        
In specific, a random phase shift vector is generated at the start of the loop containing the increment values for each of the $2N$ phase shifts. The aim of the \textit{for} loop which starts from step $7$ of the algorithm is to gradually increase the utility function value by changing the phase shifts and examining the change in the utility function when each phase shift is incremented or decremented. Specifically, the phase shift $\Omega_n$ is first incremented by its corresponding increment value $\Delta \Omega_n$; if this change does not result in an increase in the utility value, the new $\Omega_n$ is decremented by $2 \Delta \Omega_n$ which means that the original phase shift is decremented by $\Delta \Omega_n$. If the utility value still does not increase after this change, the phase shift is restored to its original value, the phase increment $\Delta \Omega_n$ is reduced by the scale factor $\alpha$ and the above procedure is repeated. When the utility value is increased, the new phase shift is accepted and the algorithm proceeds to modify the next phase shift. 
Furthermore, if the maximum number of iterations is reached without any improvement in the utility value, the algorithm gives up on $\Omega_n$, keeps it unaltered, and moves to the next phase shift.

We can control the shape of the beam by changing the value of $\epsilon$. In particular, a larger $\epsilon$ implies a more relaxed requirement on uniform
broadness while a smaller $\epsilon$ makes this requirement more strict. 
Note that if the value of $\epsilon$ is very close to zero, the utility function value might never become smaller than $\epsilon$. There is no straightforward way to determine how small $\epsilon$ can be in Algorithm~\ref{alg:epsilon-comp-pairs}. One way is to fix the number of maximum iterations and use a bisection approach for finding the smallest $\epsilon$ value for which the utility function gets below $\epsilon$ for that specific number of maximum iterations. In our simulations, we set $\epsilon$ as $2\%$ of the maximum initial sidepeak level. In other words, assuming that $\boldsymbol{\phi}^{(0)}$ is the initial RIS configuration vector, we set $\epsilon = 0.02\, F(\boldsymbol{\phi}^{(0)})$. We also set the maximum number of iterations as $L_1 = L_2 = 1000$.

We will now provide a numerical example to showcase the behavior of the algorithm. Consider a ULA-type RIS with $N = 64$ elements and the element spacing between adjacent elements being $\Delta_y = \lambda/4$. Therefore, the spacing between two successive elements of the same polarization is $\lambda/2$. The BS is a ULA with $M = 4$ dual-polarized antennas and the antenna spacing of $\Delta_{\mathrm{B}} 
= \lambda/2$. The small-scale fading of the channel between the BS and the RIS is modeled by Rician fading as 
\begin{equation}
\label{eq:Rician_channel}
  \sqrt{G_{\mathrm{B},0}(\vartheta)G_{\mathrm{R},0}(\tilde{\varphi})} \left( \sqrt{\frac{\kappa}{\kappa + 1}} \vect{H}_{\mathrm{p,LoS}} + \sqrt{\frac{1}{\kappa + 1}} \vect{H}_{\mathrm{p,NLoS}}\right),
\end{equation}
where $\kappa = 3$ is the Rician factor, and $\vect{H}_{\mathrm{p,LoS}}$ and $\vect{H}_{\mathrm{p,NLoS}}$ are the LoS and NLoS parts of the channel in polarization $\mathrm{p}$. In particular, $\vect{H}_{\mathrm{p,LoS}} = \vect{a}_{\mathrm{p}}(\Tilde{\varphi}) \vect{b}^{\Ttran}(\vartheta)$ where the AoD from the BS is $\vartheta = - \pi/3$ and the AoA to the RIS is set as $\Tilde{\varphi} = \pi/3$. 
For $\vect{H}_{\mathrm{p,NLoS}}$, we consider correlated Rayleigh fading and use the local scattering spatial correlation model with Gaussian distribution \cite[Chapter 2]{massivemimobook}.
To elaborate more, let us express the NLoS channel matrix as $\vect{H}_{\mathrm{p,NLoS}} = [\vect{h}_{\mathrm{p},1},\ldots,\vect{h}_{\mathrm{p}, M}]^{\Ttran}$. Using the correlated Rayleigh fading, this channel is modeled as $\vect{h}_{\mathrm{p},m} \sim \CN (\vect{0}_N,\vect{R})$ in which $\vect{R} \in \mathbb{C}^{N \times N}$ is the spatial correlation matrix whose $(n,n^\prime)$th entry is given by 
\begin{equation}
[\vect{R}]_{n,n^\prime} = \int e^{j\frac{2\pi}{\lambda}\left( 2\Delta_y(n - n^\prime)\sin(\tilde{\varphi} + \eta) \right)}f(\eta) d\eta,
\end{equation}where $f(\cdot)$ represents the probability density function and $\eta \sim \mathcal{N}(0,(\pi/18)^2)$. 
In Algorithm~\ref{alg:epsilon-comp-pairs}, we set $\alpha = 0.97$. 

\begin{figure}
	\centering
	\begin{subfigure}{0.495\textwidth}
		\centering
		\includegraphics[width=0.9\columnwidth]{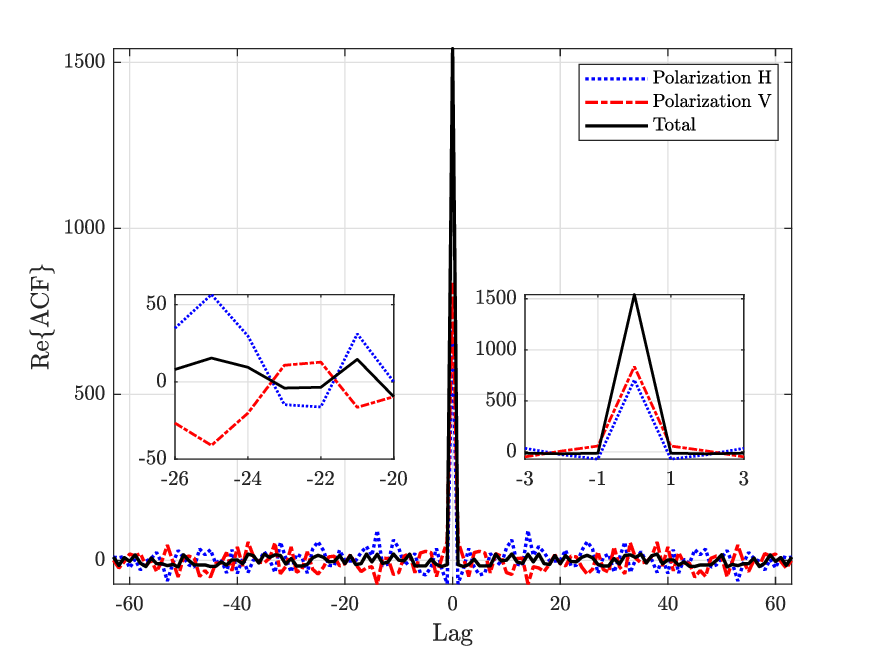}
		\caption{Auto-correlation function.}
		\label{fig:ACF_eps_comp}
	\end{subfigure}
	\begin{subfigure}{0.495\textwidth}
		\centering
		\includegraphics[width=0.9\columnwidth]{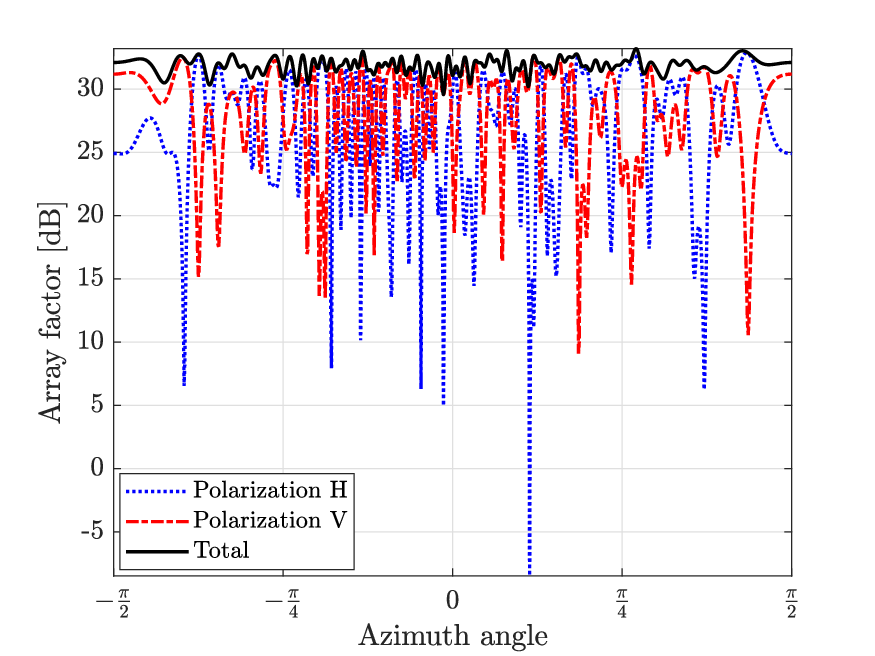}
		\caption{Power-domain array factor.}
		\label{fig:AF_eps_comp}
	\end{subfigure}
	\caption{ACF and power-domain array factor for a pair of RIS configurations obtained via the $\epsilon$-complementary algorithm.}
\label{fig:eps_comp}
\end{figure}

Fig.~\ref{fig:eps_comp} depicts the ACF and power-domain array factor when Algorithm~\ref{alg:epsilon-comp-pairs} is utilized for finding a pair of $\epsilon$-complementary RIS configurations. We can see in Fig.~\ref{fig:ACF_eps_comp} that the total ACF graph approximately resembles a Kronecker delta function with a tall peak at the origin and negligible sidepeak levels. Fig.~\ref{fig:AF_eps_comp} plots the corresponding power-domain array factor. It is observed that the per-polarization power-domain array factors exhibit substantial fluctuations over the considered angles, just as in the LoS case studied previously. However, the variation in the total array factor is rather small. 
Although the total array factor goes up and down, its value is around $32\,$dB for all observation angles. Note that one can always adjust $\epsilon$ to relax or tighten up the broad-beam condition.

\subsection{Extension from ULA to UPA}

We now consider a UPA-type RIS, in which case the power-domain array factor is a function of both azimuth and elevation angles and is given by \eqref{eq:array_factor_arbitrary}. Similar to Proposition~\ref{prop:Golay_pair}, we define matrices $\hat{\boldsymbol{\Upsilon}}_{\mathrm{H}}$ and $\hat{\boldsymbol{\Upsilon}}_{\mathrm{V}}$ where the first $N_y$ entries of $\hat{\boldsymbol{\phi}}_{\mathrm{p}}$ constitute the  first column of $\hat{\boldsymbol{\Upsilon}}_{\mathrm{p}}$, the second $N_y$ entries of $\hat{\boldsymbol{\phi}}_{\mathrm{p}}$ form the second column of $\hat{\boldsymbol{\Upsilon}}_{\mathrm{p}}$, and so forth. Then, the $\epsilon$-complementary condition for the UPA-type RIS can be expressed as 
\begin{equation}
\label{eq:epsilon_comp_UPA}
    \Big | R_{\hat{\boldsymbol{\Upsilon}}_{\mathrm{H}}}[\xi_1,\xi_2] + R_{\hat{\boldsymbol{\Upsilon}}_{\mathrm{V}}}[\xi_1,\xi_2]\Big| \leq \epsilon,~~~ \forall (\xi_1, \xi_2) \neq (0,0).
\end{equation}
We thus need to find the RIS configurations in such a way that the sum ACF is less than the threshold determined by $\epsilon$ at all points except origin. The utility function for this scenario is defined as 
\begin{equation}    F(\boldsymbol{\Upsilon}) = - \underset{(\xi_1,\xi_2) \in \mathcal{T}}{\max} \Big | R_{\hat{\boldsymbol{\Upsilon}}_{\mathrm{H}}}[\xi_1,\xi_2] + R_{\hat{\boldsymbol{\Upsilon}}_{\mathrm{V}}}[\xi_1,\xi_2]\Big|,
\end{equation}where $\boldsymbol{\Upsilon} = [\boldsymbol{\Upsilon}_{\mathrm{H}},\boldsymbol{\Upsilon}_{\mathrm{V}}]$ and the set $\mathcal{T}$ consists of all combinations of $-N_y+1 \leq \xi_1 \leq N_y -1$ and $-\tilde{N}_z+1 \leq \xi_2 \leq \Tilde{N}_z -1$ except $(\xi_1, \xi_2) = (0,0)$. Then, a problem similar to $\mathbb{P}$ can be formulated for optimizing the phase shifts of the dual-polarized RIS and Algorithm~\ref{alg:epsilon-comp-pairs} can be used to find the $\epsilon$-complementary pair of configurations for the UPA-type RIS where $\boldsymbol{\Omega}$ is modified as $\boldsymbol{\Omega} = \arg \left(\mathrm{vec}(\boldsymbol{\Upsilon}) \right)$ and $F(\boldsymbol{\phi})$ is replaced by $F(\boldsymbol{\Upsilon})$.

The complexity of Algorithm~\ref{alg:epsilon-comp-pairs} for finding a pair of $\epsilon$-complementary configurations for a UPA-type RIS grows cubically with the number of RIS elements. Specifically, the complexity of computing the 2D ACF quadratically grows with the number of elements and since we have a \textit{for} loop over the number of elements within which the 2D ACF is calculated several times, the overall complexity is in the order of $O(N^3 L_1 L_2)$. Note that we use the \texttt{xcorr2} MATLAB function for calculating the 2D ACF in the Algorithm. We can reduce the complexity of the algorithm to make it increase only quadratically with the number of elements. Specifically, according to \eqref{eq:ACF2}, the whole 2D ACF does not need to be computed each time we change one of the $2N$ phase shifts. We can only recompute the terms related to the modified phase shift in which case the complexity of computing 2D ACF will linearly grow with the number of elements and the complexity of the overall algorithm will be in the order of $O(N^2 L_1 L_2 )$.

\begin{remark}
   Since the BS and the RIS are fixed in position, the channel between them experiences minimal variation over time. As a result, there is no need to frequently update the RIS configurations. Once the RIS configurations for broad beamforming have been designed, they can be applied whenever the BS broadcasts cell-specific information.
\end{remark}

\begin{figure*}
	\centering
	\begin{subfigure}{0.33\textwidth}
		\centering
		\includegraphics[width=\columnwidth]{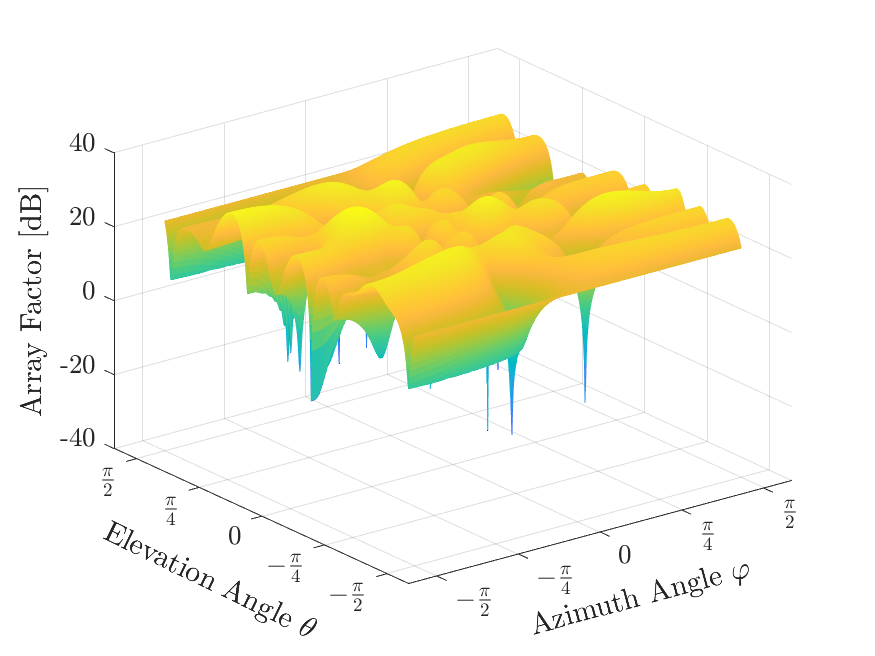}
		\caption{Polarization H}
		\label{fig:AF-eps_polH}
	\end{subfigure}%
	\begin{subfigure}{0.33\textwidth}
		\centering
		\includegraphics[width=\columnwidth]{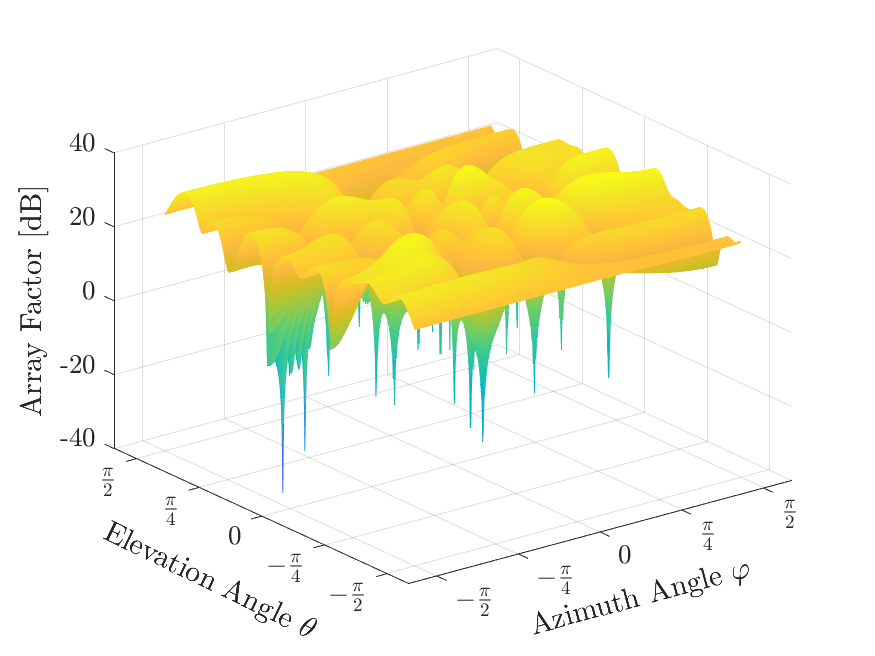}
		\caption{Polarization V}
		\label{fig:AF-eps_polV}
	\end{subfigure}%
 \begin{subfigure}{0.33\textwidth}
		\centering
		\includegraphics[width=\columnwidth]{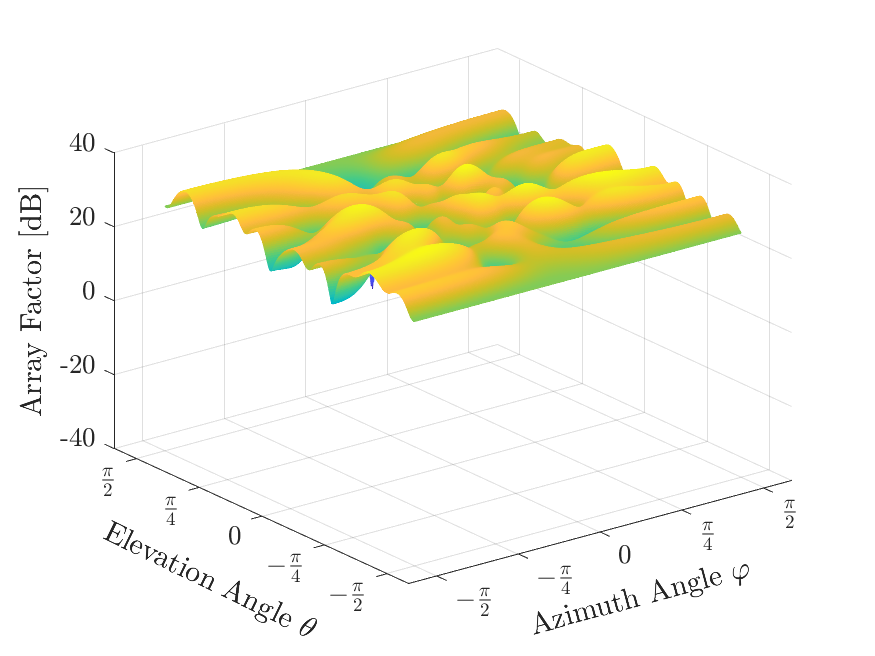}
		\caption{Total}
		\label{fig:AF-eps_tot}
	\end{subfigure}
	\caption{Illustration of the per-polarization and total power-domain array factor for a dual-polarized RIS for which the phase configurations of the two polarizations are obtained via the $\epsilon$-complementary approach. The RIS setup is the same as in Fig.~\ref{fig:AF} and a Rician fading channel is considered between the BS and the RIS.}
 \label{fig:AF_eps_UPA}
\end{figure*}

In Fig.~\ref{fig:AF_eps_UPA}, we show the power-domain array factor for a UPA-type RIS of the same setup as in Fig.~\ref{fig:AF} and $M = 4$. A Rician fading channel is considered between the BS and the RIS where all the channel parameters are similar to the ones considered in Fig.~\ref{fig:eps_comp}, the elevation AoA to the RIS is set as $\Tilde{\theta} = \pi/3 $, and the entry $(n,n^\prime)$ of the spatial correlation matrix is given by \cite{Demir2022channel}
\begin{equation}
\begin{aligned}
    [\vect{R}]_{n,n^\prime} & = \int \int e^{j\frac{2\pi}{\lambda}\left(\Delta_y (n_y-n^\prime_y)\sin(\tilde{\varphi}+\eta_1)\cos(\tilde{\theta} + \eta_2) \right) }  \\
    & \times e^{j\frac{2\pi}{\lambda}\left(2\Delta_z(n_z -n^\prime_z) \sin(\tilde{\theta} + \eta_2)\right)} f(\eta_1) f(\eta_2) d\eta_1 d\eta_2 
    \end{aligned}
\end{equation}where    $\eta_1, \eta_2 \sim \mathcal{N}\left(0,(\pi/18)^2\right)$, $n_y (n^\prime_y) = \mod \left(n (n^\prime)-1,N_y\right) +1$,  and $n_z(n^\prime_z) = \left\lceil n(n^\prime)/N_y \right\rceil$.
We can see that the total power-domain array factor is not perfectly plane as it was in Fig.~\ref{fig:AF} and there exist ripples due to the non-zero sidepeaks in the total ACF. However, it has a relatively flat and smooth shape because the amount of fluctuations in the total array factor is guaranteed to be limited. 

\section{Numerical Results}
\label{sec:results}
In this section, we present our numerical results to further demonstrate and evaluate the performance of the proposed broad beamforming approaches. 

Assuming a LoS channel between the BS and the RIS, we consider a $16 \times 16$ RIS with the same setup and configurations used in Fig.~\ref{fig:AF}. We show in Fig.~\ref{fig:radiation-pattern} the total power radiation pattern of the dual-polarized RIS over $\varphi \in [-\pi/2,\pi/2]$ and $\theta \in [-\pi/2, \pi/2]$. The RIS power radiation pattern is given by $G_{\mathrm{R},0}(\varphi,\theta)A(\varphi,\theta)$ and the radiation pattern of one RIS element is modeled by the 3GPP antenna gain model \cite[Table~7.1-1]{3gpp} as
\begin{align}
    G_{\mathrm{R},0}(\varphi,\theta) =&\; 8 -\min \bigg[ \min \Big[12 \left( \frac{\varphi - \varphi_0}{\Delta \varphi}\right)^2,30 \Big] \notag \\ & + \min \Big[12 \left( \frac{\theta - \theta_0}{\Delta \theta}\right)^2,30 \Big], 30 \bigg]~~~\mathrm{[dBi]},
\end{align}with $\varphi_0 = \theta_0 = 0$ and $\Delta \varphi = \Delta \theta = \pi/2$. Fig.~\ref{fig:radiation-pattern}  shows that the RIS whose configuration matrices form a Golay complementary array pair preserves the broad radiation pattern of a single RIS element, while shifting it upward. This endorses the effectiveness of the proposed approach based on Golay complementary pairs for producing uniformly broad beams from the RIS when the channel between the BS and the RIS is purely LoS. In other words, we preserve the radiation pattern but benefit from the power gain by having many elements that reflect signals from the BS.

\begin{figure}
    \centering
    \includegraphics[width=0.9\columnwidth]{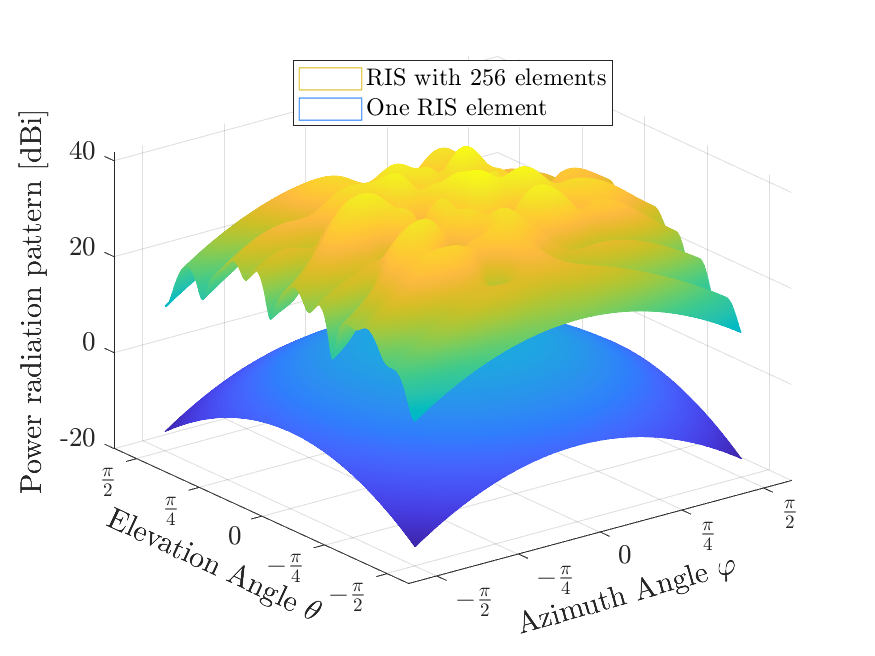}
    \caption{Illustration of the power radiation pattern for a $256$-element dual-polarized RIS when its phase configuration matrices are obtained via the $\epsilon$-complementary approach.  }
    \label{fig:radiation-pattern_eps_complementary}
\end{figure}

In Fig.~\ref{fig:radiation-pattern_eps_complementary}, we consider a Rician fading channel between the BS and the RIS where the parameters are the same as those of Fig.~\ref{fig:AF_eps_UPA}. The figure shows the power radiation pattern of the dual-polarized RIS where its phase shift configurations are obtained via the $\epsilon$-complementary approach outlined in Algorithm~\ref{alg:epsilon-comp-pairs} with  $\alpha = 0.97$.  It can be observed that there exist small fluctuations in the power radiation pattern because the requirement of having zero sidepeaks in the sum ACF is relaxed in the $\epsilon$-complementary approach. Nevertheless, the beam radiated by the RIS has a broad shape, implying that the RIS can cover all the users who reside in different azimuth and elevation angles almost uniformly.

We now proceed to evaluate the end-user performance of the proposed broad beamforming schemes. We assume $K = 1000$ users are distributed around the RIS, where the path-loss for the channel between the RIS and an arbitrary user is modeled as 
\begin{equation}
    \beta  = -37.5 - 22 \log_{10} \big(d/1~ \mathrm{m}\big) ~~[\mathrm{dB}],
\end{equation}where $d$ represents the distance between the RIS and the user. 
The path-loss from the BS to the RIS, $\Tilde{\beta}$, is modeled in a similar way with $\Tilde{d}$ being the distance between the BS and the RIS. The setup for the BS and the RIS (i.e., the number of BS antennas, number of RIS elements, AoD from the BS, AoA at the RIS, etc.) is the same as before. Furthermore, the distance from the BS to the RIS is set as $\Tilde{d}= 50\,$m, the BS transmit power is set as  $P_{\mathrm{T}} = 30\,$dBm, and the noise power is set to be $\sigma^2 = -110\,$dBm. 
The users are uniformly distributed around the RIS such that $d \sim \mathcal{U}[50\,\mathrm{m},80\,\mathrm{m}]$,  
$\varphi \sim \mathcal{U}[-\pi/3,\pi/3]$, and $\theta \sim \mathcal{U}[-\pi/6,\pi/6]$. 
 
 We compare the SE offered by the proposed designs with the SE obtained by three benchmark approaches:

 \begin{figure}
    \centering
    \includegraphics[width=0.9\columnwidth]{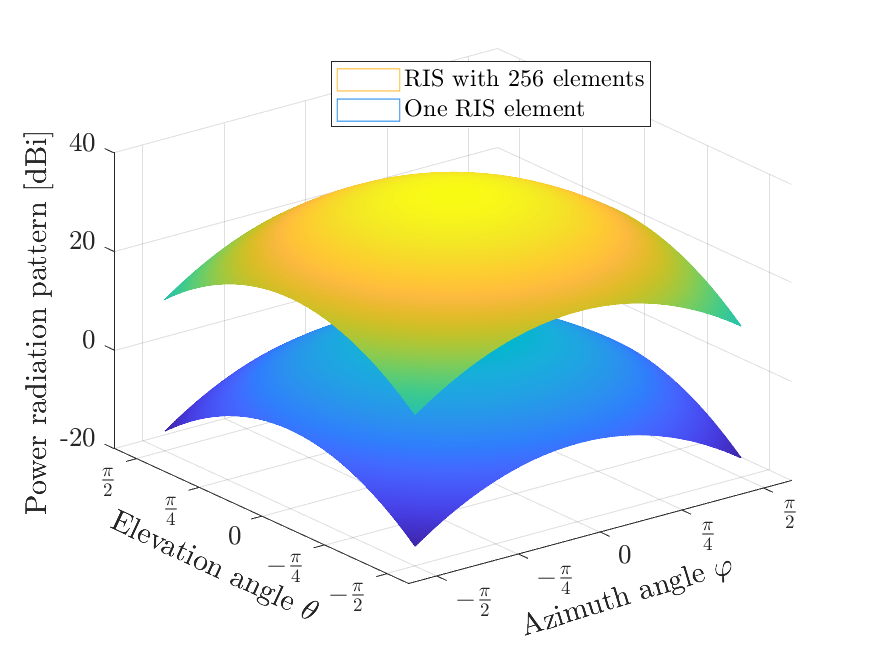}
    \caption{Illustration of the power radiation pattern for a $256$-element dual-polarized RIS when its phase configuration matrices are a Golay complementary array pair.  }
    \label{fig:radiation-pattern}
\end{figure}

\begin{itemize}

\item \textbf{DP-MaxSum}: Assuming the availability of perfect CSI, the configurations of the dual-polarized RIS are designed to maximize the sum received power at the given set of users.
    \item \textbf{DP-MaxMin}: The interval $[-\pi/2,\pi/2]$ is split into $100$ uniformly-spaced angles in both azimuth and elevation domains. The configurations of the dual-polarized RIS are designed to maximize the minimum received power in these $10000$ angular directions. In such a case, the beam radiated by the RIS resembles a broad beam.
   \item   \textbf{UP-MaxMin}: This approach is similar to DP-MaxMin, except that  here, a uni-polarized RIS is utilized with its number of elements being equal to the total number of elements across both polarizations in the dual-polarized RIS. Similar to the previous benchmark, the area in front of the RIS is divided into $100$ uniformly-spaced azimuth and elevation angles and the minimum received power is maximized.
\end{itemize}

The SE is computed as

\begin{equation}
\label{eq:SE}
  \mathrm{SE} = \log_2\left(1+ \mathrm{SNR}\right)~[\mathrm{bps/Hz}],   
\end{equation} where $\mathrm{SNR}$ is given in \eqref{eq:SNR_LoS} for the case where the BS-RIS channel is LoS, while \eqref{eq:SNR_arbitrary} represents the SNR for the case with arbitrary channel between the BS and the RIS. 

\begin{figure}
    \centering
    \includegraphics[width=\columnwidth]{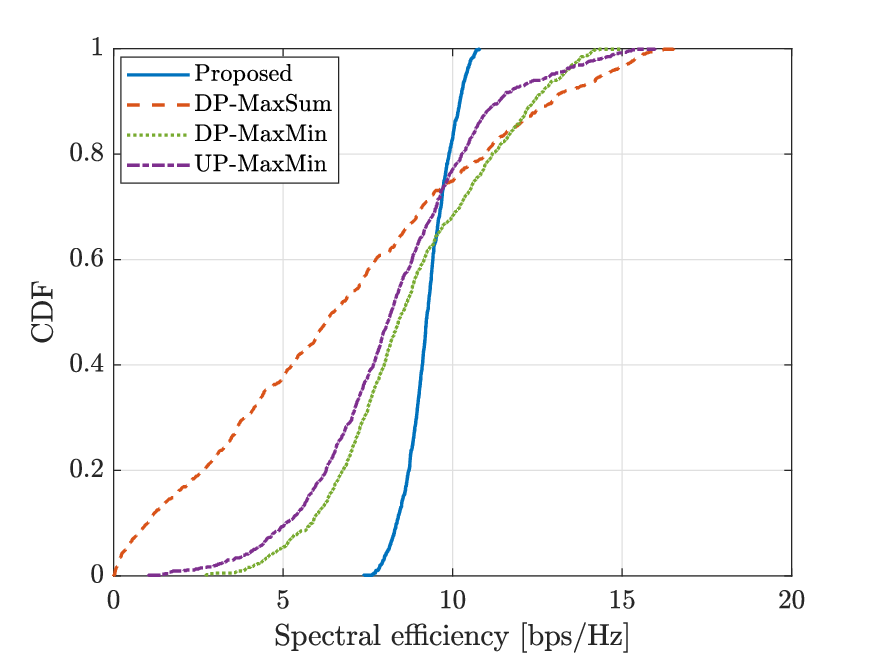}
    \caption{CDF of SE when an LoS channel is considered between the BS and the RIS. }
    \label{fig:SE_Golay}
\end{figure}

Fig.~\ref{fig:SE_Golay} shows the cumulative distribution function (CDF) of SE for the proposed and benchmark schemes when the channel between the BS and the RIS is assumed to be purely LoS. The proposed scheme provides more than $73\%$ of the users with a higher SE than that provided by DP-MaxSum and UP-MaxMin approaches. Furthermore, more than $64\%$ of the users get a higher SE with the proposed scheme as compared to the DP-MaxMin approach. More importantly, with all the benchmark approaches, many users are provided with very low SEs.
This is undesirable in the broadcasting scenarios considered in this paper because the minimum SE restricts how much data can be broadcast as all users must be able to decode. When using the benchmark methods, we must either broadcast at a low SE or drop a large fraction of the users from service to achieve a decent minimum SE among the remaining ones. By contrast, the proposed method enables broadcasting to all users at a relatively high SE.

\begin{figure}
    \centering
    \includegraphics[width=\columnwidth]{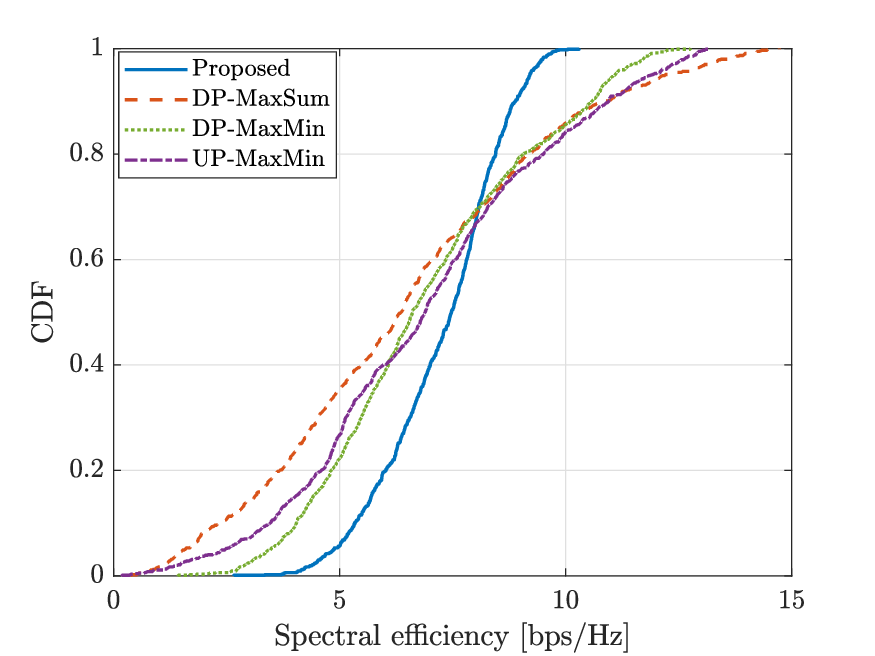}
    \caption{CDF of SE when a Rician fading channel is considered between the BS and the RIS. } 
    \label{fig:SE_eps_comp}
\end{figure}

Next, Fig.~\ref{fig:SE_eps_comp} presents the CDF of SE considering a Rician fading channel between the BS and the RIS with the same parameters used in Figs.~\ref{fig:AF_eps_UPA} and \ref{fig:radiation-pattern_eps_complementary}. 
The observations are similar to those of Fig.~\ref{fig:SE_Golay}. More than $68\%$ of the users are provided with a higher SE with the proposed approach as compared to the benchmarks. The SE value has less fluctuations over the set of users with the proposed approach, which is appealing for broadcasting scenarios. 

\begin{figure}
    \centering
    \includegraphics[width=\columnwidth]{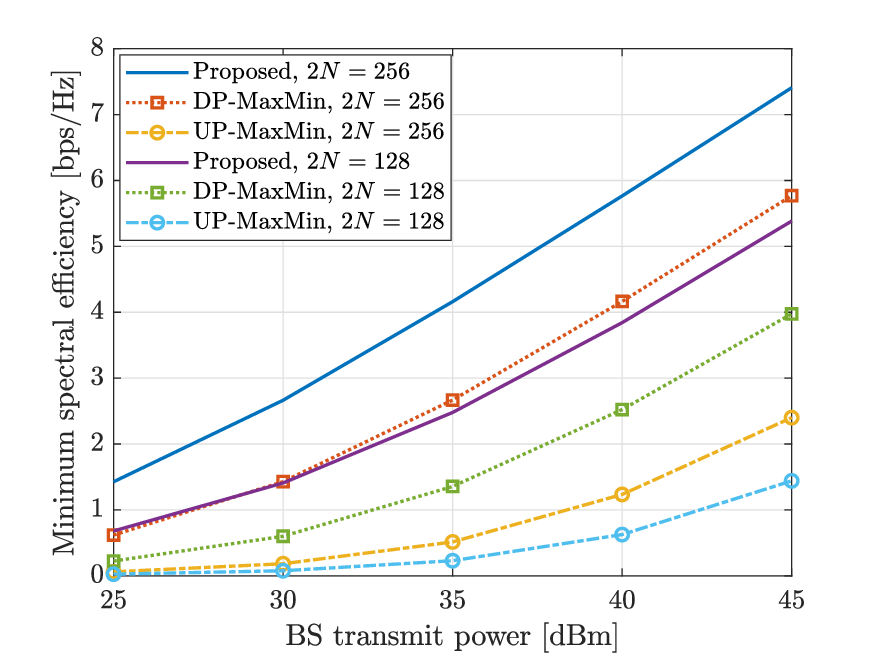}
    \caption{Minimum SE achieved by the proposed scheme, DP-MaxMin, and UP-MaxMin, in case of a Rician fading channel. } 
    \label{fig:minSE}
\end{figure}

An interesting observation can be made by comparing the minimum SEs provided by DP-MaxMin and UP-MaxMin approaches. Figs. \ref{fig:SE_Golay} and \ref{fig:SE_eps_comp} demonstrate that the minimum SE provided by the DP-MaxMin scheme is superior to that provided by the UP-MaxMin approach although they are both designed following the same strategy, which is maximizing the minimum SNR across all angular directions. 
To better illustrate this, we show in Fig.~\ref{fig:minSE} the minimum SE achieved by our proposed scheme, DP-MaxMin, and UP-MaxMin as a function of BS transmit power, in case of a Rician fading BS-RIS channel. One can notice that the minimum SE offered by the DP-MaxMin scheme is higher than that provided by the UP-MaxMin scheme, and our proposed scheme outperforms both. The higher SE of DP-MaxMin compared to UP-MaxMin is due to the fact that the former takes advantage of the polarization degree of freedom which allows the users to receive signals over two polarizations. The gap between the minimum SEs provided by DP-MaxMin and UP-MaxMin is greater when RIS has more elements as observed by comparing the cases with $256$ and $128$ RIS elements.

\section{Conclusions and Future Outlook}
\label{sec:conclusions}
In this paper, we have proposed methods for achieving broad radiated beams from a dual-polarized RIS, which is desirable when the RIS supports the BS in broadcasting of system information. The key is to design the RIS configurations of the two polarizations such that the beams radiated from different polarizations complement each other and create an overall broad radiation pattern. This can be accomplished when the power spectra of the RIS configuration pair add up to a constant.  
We first considered a LoS channel between the BS and the RIS. In this case, a uniformly broad beam can be achieved by letting the RIS configuration matrices form a Golay complementary array pair. Then, we investigated a general channel scenario between the BS and the RIS. By relaxing the requirement of having a uniformly broad beam and admitting some level of imperfection in the sum power spectrum, we presented the $\epsilon$-complementary algorithm which finds a pair of RIS configurations for producing a practically broad beam through stochastic optimization. The proposed algorithms can be utilized for RIS-assisted cell-specific beamforming from the BS to users who reside at unknown locations. 

This paper has investigated broad beam design for dual-polarized RIS-assisted systems, aiming to uniformly cover all angular directions around the RIS. 
In practice, the BS may have non-negligible channels to users in certain parts of the intended coverage area. If this information is available (e.g., in the form of a digital twin), the beam design at the RIS can be adjusted so it spreads less power in directions where the BS already provides decent coverage and more power in directions that are not adequately covered by the BS. Therefore, designing the per-polarization RIS configurations for producing semi-broad beams that are fine-tuned to a particular deployment area would be an important direction for future research.

\bibliographystyle{IEEEtran}
\bibliography{refs}
\end{document}